\documentclass[a4paper,prd,preprintnumbers,twocolumn,superscriptaddress,nofootinbib,amsmath,amssymb]{revtex4-1}

\usepackage[dvips]{graphics}
\usepackage{color,hyperref}
\usepackage{times}
\usepackage{mathrsfs}
\hypersetup{colorlinks=true,linkcolor=blue,citecolor=red,filecolor=magenta,urlcolor=blue}

\usepackage{enumitem}
\usepackage{graphicx,subfigure}
\usepackage{dcolumn}
\usepackage{bm}
\usepackage{color}

\begin{document}
\title{Circular particle motion around Schwarzschild-MOG black hole} 

\author{Madina Boboqambarova}\email{asmik96@astrin.uz}\affiliation{Ulugh Beg Astronomical Institute, Astronomy St. 33, Tashkent 100052, Uzbekistan}

\author{Bobur Turimov}\email{bturimov@astrin.uz}
\affiliation{Ulugh Beg Astronomical Institute, Astronomy St. 33, Tashkent 100052, Uzbekistan}\affiliation{Akfa University, Milliy Bog St. 264, Tashkent 111221, Uzbekistan}\affiliation{Institute of Nuclear Physics, Ulugbek 1, Tashkent 100214, Uzbekistan}
\affiliation{Ajou University in Tashkent, Asalobod St. 113, Tashkent 100204, Uzbekistan}\affiliation{Webster University in Tashkent, Alisher Navoiy St. 13, Tashkent 100011, Uzbekistan}

\author{Ahmadjon Abdujabbarov}\email{ahmadjon@astrin.uz}
\affiliation{Ulugh Beg Astronomical Institute, Astronomy St. 33, Tashkent 100052, Uzbekistan}
\affiliation{Institute of Nuclear Physics, Ulugbek 1, Tashkent 100214, Uzbekistan}
\affiliation{Shanghai Astronomical Observatory, 80 Nandan Road, Shanghai 200030, P. R. China}
\affiliation{National University of Uzbekistan, Tashkent 100174, Uzbekistan}
\affiliation{Institute of Fundamental and Applied Research, National Research University TIIAME, Kori Niyoziy 39, Tashkent 100000, Uzbekistan}
\affiliation{Webster University in Tashkent, Alisher Navoiy St. 13, Tashkent 100011, Uzbekistan} 
\date{\today}
\begin{abstract}
In this paper, we present an analysis of the circular motion of test particles around a Schwarzschild-MOG black hole. Initially, our focus lies on studying the shadow cast by the spherically symmetric black hole within the framework of MOG gravity. Notably, we observe that the presence of MOG influences both the photon sphere and the black hole's shadow, causing them to increase in size. Furthermore, our research reveals that the characteristic radii of massive particles in circular orbits around the Schwarzschild-MOG black hole, specifically the innermost stable circular orbits (ISCO) and marginally bound orbits, are greater than those observed in the Schwarzschild metric alone. Additionally, we examine the electromagnetic field structure when a black hole is subjected to an external uniform magnetic field. Our findings demonstrate that in the vicinity of the Schwarzschild-MOG black hole, the magnetic field exhibits non-uniform behavior, with field lines becoming more densely distributed. Lastly, we delve into the motion of charged particles around the Schwarzschild-MOG black hole in the presence of an external magnetic field. Our investigation highlights that the ISCO position for charged particles is consistently less than that for neutral particles, indicating a significant distinction between the two scenarios.
\end{abstract}
\maketitle
\section{Introduction}

In recent years, numerous independent experiments and observations have provided compelling evidence for the existence of dominant forms of matter in the Universe: dark energy and dark matter~\cite{Bertone2018Nature, Bertone2018RMP}. However, a comprehensive theoretical framework to explain the nature and behavior of these forms of matter is still lacking. The standard gravity theory, general relativity, which is a classical field theory, also faces fundamental challenges such as singularities in certain solutions and its inconsistency with quantum field theory. To address these issues, modifications to general relativity or alternative theories of gravity have been proposed. There is a wide range of attempts to modify and generalize Einstein's theory of gravity, one of which is the scalar-tensor-vector gravity (STVG) theory proposed by Moffat~\cite{Moffat2006CQG}. This theory introduces scalar and massive vector fields and aims to construct a unified theory of gravity, commonly referred to as the MOG theory.

MOG theory has been extensively studied, and several solutions describing compact objects have been derived~\cite{Moffat13, Moffat15}. Specifically, detailed investigations have been conducted on the black hole shadow~\cite{Moffat15a, Moffat2015EPJC, Moffat2020PRD}, quasi-normal modes, and superadiance of scattered waves by black holes~\cite{Manfredi2018PLB, Wondrak18}, as well as gravitational lensing~\cite{Moffat09, Rahvar2019MNRAS}. Spherical black hole solutions within this theory have been obtained in~\cite{Moffat15}. The influence of STVG on the motion of the S2 star around the supermassive black hole at the center of the Milky Way has been investigated in~\cite{DellaMonica2022MNRAS, 2022Univ, Turimov2022MNRAS}. Furthermore, the motion of magnetized particles around MOG black holes in the presence of an external magnetic field has been studied in~\cite{Haydarov2020EPJC}.

The choice of a particular gravity theory determines the properties of spacetime and the corresponding solutions describing astrophysical compact objects. Among various ways to test gravity and the corresponding spacetime solutions around compact objects, the motion of test particles is one of the most efficient probes of gravity~(see, e.g.,~\cite{Jawad16, Hussain15, Babar16, Banados09, Majeed17, Zakria15, Brevik19}). Additionally, the presence of an external electromagnetic field around astrophysical compact objects affects the motion of charged or magnetically charged particles~\cite{Chen16, Hashimoto17, Dalui19, Han08, Moura00}. Detailed studies of electromagnetic field solutions around black holes and testing them through the motion of charged particles can be found in references~\cite{Wald74, Aliev86, Aliev89, Aliev02, Frolov11, Frolov12, Benavides-Gallego18, Shaymatov18, Stuchlik14a, Abdujabbarov10, Abdujabbarov11a, Abdujabbarov11, Abdujabbarov08, Karas12a, Stuchlik16, Kovar10, Kovar14, Kolos17, Pulat2020PhRvDMOG, Rayimbaev2019IJMPCS, Rayimbaev2020MPLA, Rayimbaev2019IJMPD}. Moreover, the presence of an electromagnetic field around black holes necessitates the study of particles with nonzero magnetic moments~\cite{deFelice, deFelice2004, Rayimbaev16, Juraeva2021EPJC, Rayimbaev2021NuPhB, Toshmatov15d, Abdujabbarov14, Rahimov11a, Rahimov11, Haydarov20, Abdujabbarov2020PDU, Narzilloev2020PhRvDstringy, Rayimbaev2020PhRvD, TurimovPhysRevD2020, MorozovaV2014PhRvD, Vrba2020PhRvD, Vrba2019EPJC}.

The recent observation of the image of the central supermassive black hole in M87 has sparked increased interest in studying the shadows of black holes~\cite{EventHorizonTelescope:2019dse}. The data obtained by the Event Horizon Telescope (EHT) collaboration has been utilized to constrain various parameters of black hole solutions within different gravity models~\cite{Ghasemi-Nodehi:2020oiz, Jusufi:2020odz, Liu:2020ola, Jusufi:2019caq, Afrin2021a, Atamurotov2013a}. However, the properties of black hole shadows have been studied theoretically by different authors even before the EHT observation~\cite{Bambi:2014mla, Abdujabbarov2013a, Far:2016c}. In this work, we aim to investigate the shadow of a spherically symmetric black hole described by the Schwarzschild-MOG solution.

The paper is organized as follows: In Section~\ref{Sec:Spacetime}, we delve into the study of geodesic motion, focusing on the photon sphere and observable shadow. We also analyze the circular motion of test massive particles. Section~\ref{Sec:ChargedParticleMotion} is dedicated to the motion of charged particles, specifically investigating their behavior in the Schwarzschild-MOG black hole spacetime. Finally, we summarize our findings in Section~\ref{Sec:Summary}. Throughout the paper, we adopt the metric signature (-, +, +, +) for the spacetime metric and use a system of units where the gravitational constant and the speed of light are set to 1: $G_N = 1 = c$. Greek indices range from 0 to 3, while Latin indices range from 1 to 3.

\section{Geodesic motion\label{Sec:Spacetime}}

In Boyer–Lindquist coordinates $x^\alpha=(t,r,\theta,\phi)$, the spacetime around the Schwarzschild-MOG is given by the metric~\cite{Moffat15}:
\begin{align}\label{metric}
ds^2 =-fdt^2+\frac{dr^2}{f}+r^2\left(d\theta^2+\sin^2\theta d\phi^2\right)\ ,
\end{align}
with
\begin{align}\label{f}
f=1-\frac{2M(1+\alpha)}{r}+\frac{M^2\alpha(1+\alpha)}{r^2}\ , 
\end{align}
where, $M$ is the total mass of the black hole, $\alpha$ is MOG parameter. The horizon of the black hole is found as largest root of the following equation $f=0$, or
\begin{align}
r_+=M\left(1+\alpha+\sqrt{1+\alpha}\right)\ .  
\end{align}
which is always greater than the Schwarzschild radius corresponding to the positive values of $\alpha$ parameter.

The equation of motion for a test particle with specific energy ${\cal E}$ and specific angular momentum ${\cal L}$ can be expressed as follows:

\begin{align}
\dot{t} &= \frac{{\cal E}}{f}, \quad \dot{\phi} = \frac{{\cal L}^2}{r^2}, \\
\dot{r}^2 &= {\cal E}^2 - f\left(\epsilon + \frac{{\cal L}^2}{r^2}\right),
\end{align}

where the dot denotes differentiation with respect to an affine parameter along the particle's trajectory. The parameter $\epsilon$ determines whether the geodesic is null ($\epsilon = 0$) or timelike ($\epsilon = 1$). Later, we will consider the motion of both massless and massive particles separately.

It is worth noting that geodesic motion around the Schwarzschild-MOG black hole has been previously studied in~\cite{Sharif2018JETP}. However, in that paper, it is assumed that the parameter $\alpha$ is arbitrarily chosen and compared to the ratio of the gravitational constants in MOG and Newtonian theories, $G_N/G$. It is important to keep in mind that, according to references~\cite{Moffat2013MNRAS, Moffat15}, the parameter $\alpha$ is defined as $\alpha = (G - G_N)/G_N$. From this perspective, we can simply use the metric with the lapse function given in (\ref{f}). Consequently, our results may differ from those presented in Ref.~\cite{Sharif2018JETP}.

\subsection{Shadow as capture cross section of photon by black hole}

In order to test the MOG theory using the analysis of the black hole shadow, we need to determine the critical value of the impact parameter for photons, defined as the ratio of angular momentum to energy, given by $b = \frac{{\cal L}}{{\cal E}}$. Considering the motion of photons corresponding to null geodesics with $\epsilon = 0$, we can write the equation for radial motion as follows:

\begin{align}
\frac{{\dot{r}}^2}{{\cal E}^2} = 1 - f\frac{b^2}{r^2}.
\end{align}

By imposing the conditions ${\dot{r}} = {\ddot{r}} = 0$, we can determine the radius of the photon sphere and the critical value of the impact parameter as follows \cite{Moffat2015EPJC}:

\begin{align}
\frac{r_{\rm ph}}{M} &= \frac{1}{2}\left(3 + 3\alpha + \sqrt{(1+\alpha)(9+\alpha)}\right), \\
\frac{b_0}{M} &= \frac{\sqrt{2\alpha} \left(3 + 3\alpha + \sqrt{(1+\alpha)(9+\alpha)}\right)}{\sqrt{\alpha-3+\sqrt{(1+\alpha)(9+\alpha)}}}.
\end{align}

One can easily check that absence of $\alpha$ parameter above expressions will be $r_{\rm ph}=3M$ and $b_0=3\sqrt{3}M$ corresponding to the photon-sphere and critical impact parameter in the Schwarzschild space.

Capture cross section of photon by the black hole can be determined as, $\sigma=\pi b_0^2$ and in the Schwarzschild-MOG metric it takes a form:
\begin{align}
\sigma=\frac{2\pi\alpha M^2 \left(3+3\alpha+\sqrt{(1+\alpha)(9+\alpha)}\right)^2 }{\alpha-3+\sqrt{(1+\alpha)(9+\alpha)}} \ ,   
\end{align}
while in the Schwarzschild one it reduces to $\sigma=27\pi M^2\simeq 84.823M^2$. To have an idea about size of the black hole's shadow one can estimate cross section for the supermassive black hole (SMBH) with a million solar mass in the form:
\begin{align}
\sigma\simeq 1.576\times 10^{16} \left(\frac{M}{10^6M_\odot}\right)^2{\rm km}^2\ .    
\end{align}
Note that the large size of the shadow does not necessarily imply a large angular size, as it depends on the distance between the observer and the supermassive black hole.

\begin{figure*}
\includegraphics[width=\hsize]{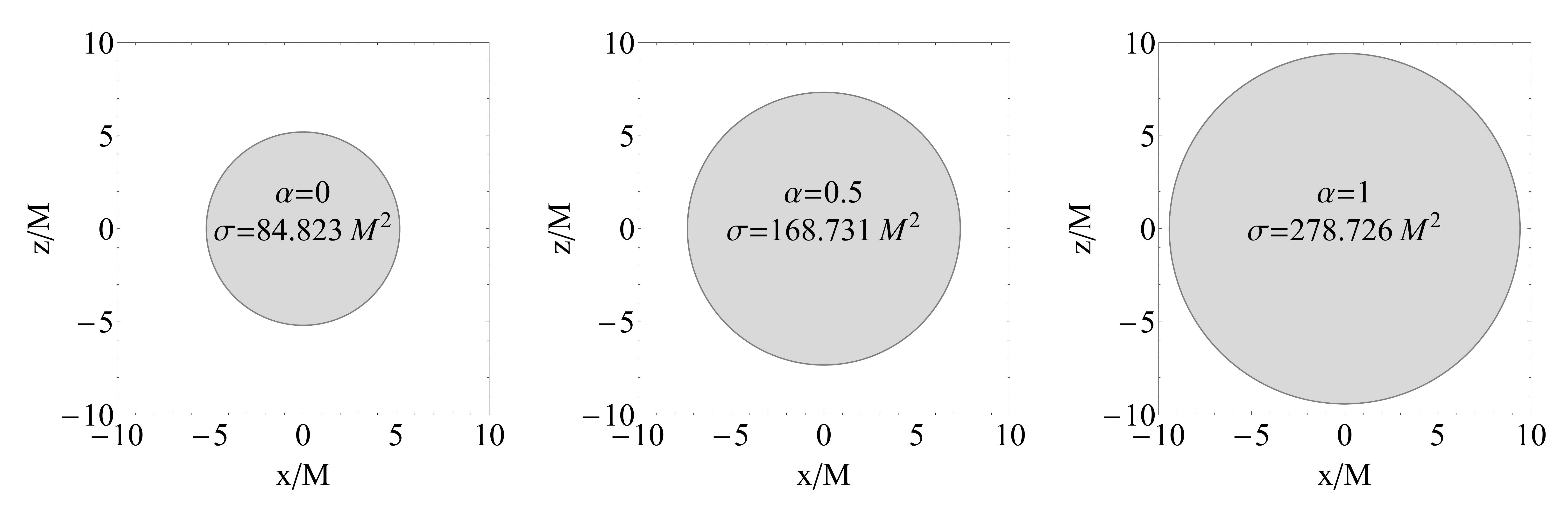}
\caption{Capture cross section of photon by the black hole (shadow of the black hole) in the (x-z) plane for the different values of $\alpha$ parameter.\label{Shadow}}
\end{figure*}

For a more detailed analysis, we can present the results graphically. Figure~\ref{Shadow} depicts the capture cross section of photons by the black hole for different values of the $\alpha$ parameter. Our analysis reveals that the capture cross section in the Schwarzschild-MOG spacetime is larger than that obtained in general relativity, which is consistent with the findings in Ref. \cite{Moffat2015EPJC}.

\subsection{The circular orbits}

In order to study a particular class of orbits for massive particles, namely the innermost stable circular orbits (ISCO) and marginally bound orbits (MBO), which are crucial for probing modified gravity in the vicinity of a black hole, we consider the particle motion with $\epsilon=1$. In this case, the radial equation for a massive particle can be written as follows:

\begin{align}
\dot{r}^2 = {\cal E}^2 - f\left(1 + \frac{{\cal L}^2}{r^2}\right).
\end{align}

Using the conditions ${\dot r}={\ddot r}=0$  corresponding to the case of absence of radial motion and radial acceleration of particle, the critical specific energy and specific angular momentum can be obtained as 
\begin{align}
&{\cal E}^2=\frac{r^2f^2}{r^2-3Mr(1+\alpha)+2M^2\alpha(1+\alpha)}\ ,\\& {\cal L}^2=\frac{Mr^2(r-\alpha  M)(1+\alpha)}{r^2-3Mr(1+\alpha)+2M^2\alpha(1+\alpha)}\ .   
\end{align}
The stationary point of these functions are responsible for the ISCO position for massive particle, which takes a form:  
\begin{align}
r_{\rm ISCO}/M=2(1+\alpha)+X+\frac{(4+\alpha)(1+\alpha)}{X}\ ,  
\end{align}
where 
$$X=\sqrt[3]{(1+\alpha)\left(8+7\alpha+\alpha^2+\alpha\sqrt{\alpha +5}\right)}$$. 
One can now easily check that switching off MOG parameter, i.e. $\alpha=0$, the ISCO position reduces $r_{\rm ISCO}=6M$ which coincides with the value of ISCO in the Schwarzschild space.  

The radius of the marginally bound orbit (MBO) can be found from the condition, ${\cal E}^2=1$, which leads the following polynomial equation:
\begin{align}\label{eqmarg} 
r^3-4(1+\alpha)Mr^2+4\alpha(1+\alpha)M^2r-\alpha^2 (1+\alpha)M^3=0\ .
\end{align}
The real root of the equation (\ref{eqmarg}) represents the MBO radius for test particle. When $\alpha=0$ one can obtain that $r_{\rm MBO}=4M$ is the MBO radius in the Schwarzschild space. 

\begin{figure}
\includegraphics[width=\hsize]{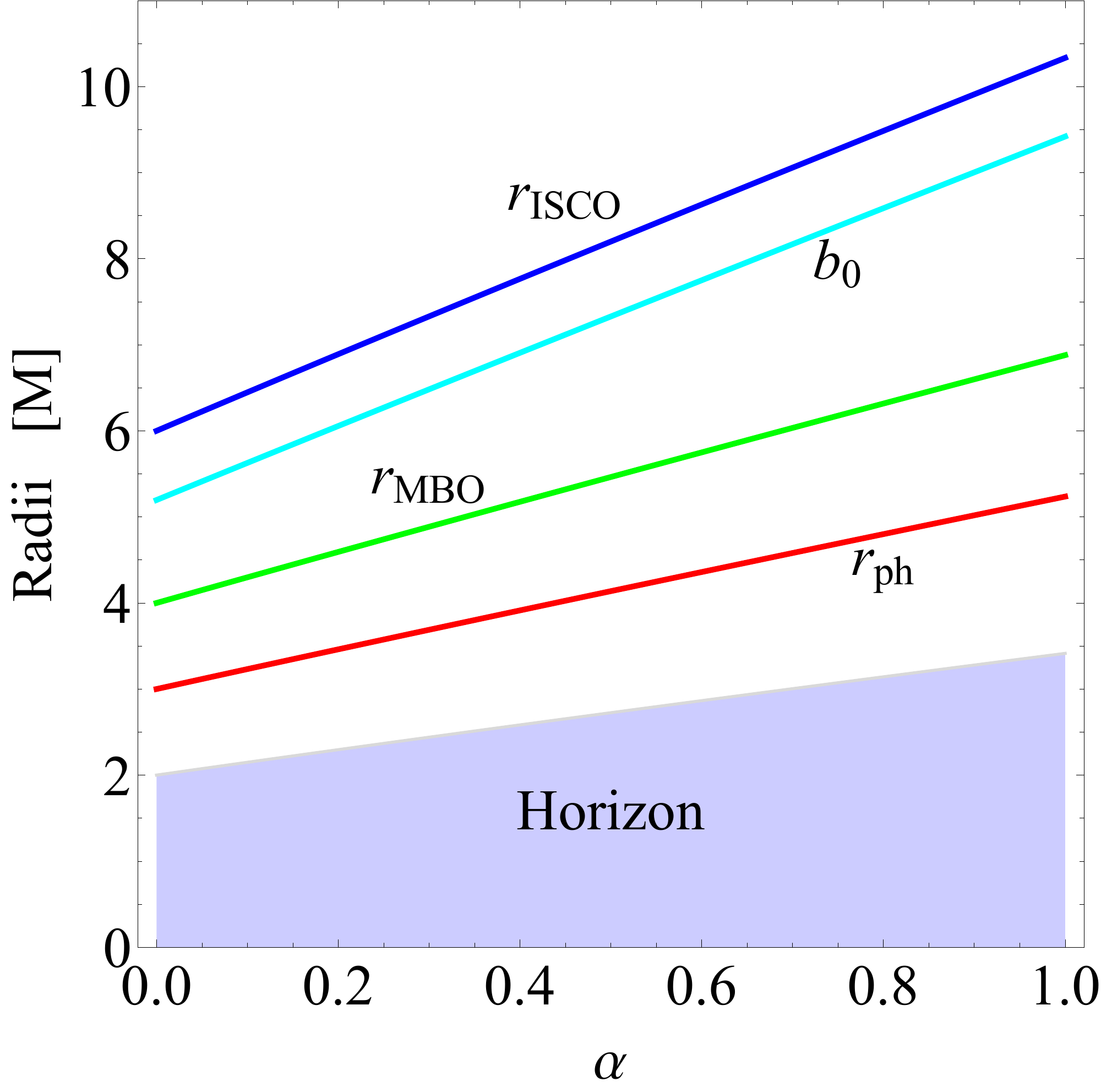}
\caption{The characteristic radii, namely, the horizon, photon-sphere, impact parameter, and ISCO position as the functions of $\alpha$ parameter.\label{Radii}}
\end{figure}

Figure~\ref{Radii} shows the dependence of the characteristic radii from $\alpha$ parameter. Taking into account all facts above, one can conclude that the characteristic radii, namely, the horizon, photon-sphere, MBO and ISCO radii for test particle get large due to the presence of $\alpha$ parameter. 

From an astrophysical perspective, the extraction of energy from a black hole is an intriguing problem. One simple model for accreting matter onto a black hole was proposed by Thorne \cite{Thorne1974ApJ}. According to this model, matter flows onto the black hole between the innermost stable orbit and the outer horizon, and the energy efficiency is related to the binding energy of the last stable orbit. The energy efficiency of a test particle can be determined as $\eta = 1 - {\cal E}_{\rm ISCO}$, where ${\cal E}_{\rm ISCO}$ is the specific energy of the particle at the ISCO position, which depends only on the parameter $\alpha$. We won't present the explicit expression for $\eta$ here, but we can determine the minimum and maximum values of the energy efficiency as follows:

\begin{align}
\lim_{{\alpha}\to0}\eta &= 1 - \frac{2\sqrt{2}}{3} \simeq 0.0572~~~(5.72\%), \\
\lim_{{\alpha}\to\infty}\eta &= 1 - \frac{3}{4}\sqrt{\frac{3}{2}} \simeq 0.0814~~~(8.14\%).
\end{align}

These limits represent the minimum and maximum values of the energy efficiency, respectively.

\begin{figure}
\includegraphics[width=\hsize]{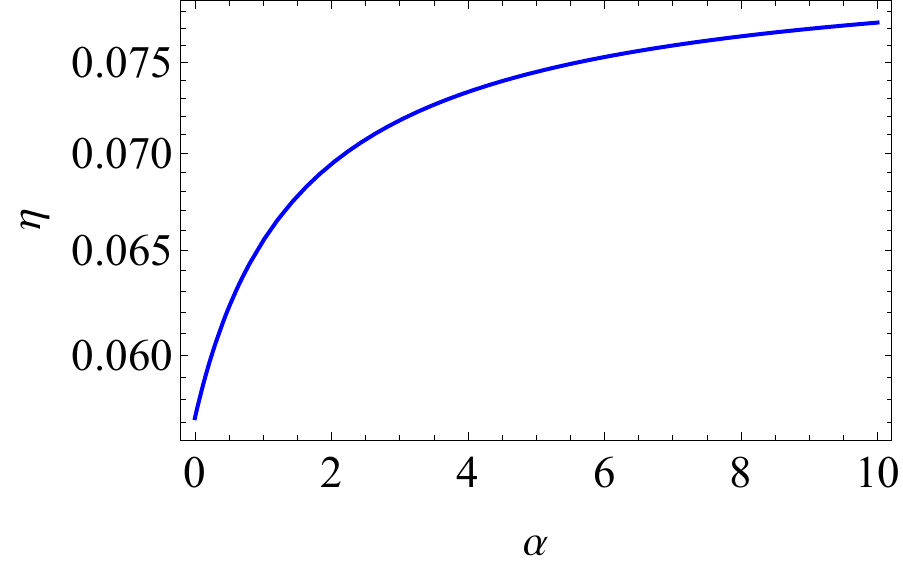}
\caption{The energy efficiency of test particle orbiting around the Schwarzschild-MOG black hole at the ISCO position.\label{eta}}
\end{figure}

The dependence of the energy efficiency from $\alpha$ parameter is illustrated in Fig.~\ref{eta}. It shows that the energy efficiency of test particle in the Schwarzschild-MOG spacetime is greater than one seen in general relativity. 

\subsection{Orbital velocities}

It is also interesting to study orbital angular and linear velocity of test particle around the black hole in order to test the theory of MOG. The angular velocity measured by a distant observer or so-called Keplerian frequency can be determined as 
\begin{align}\label{W}
\Omega=\sqrt{-\frac{\partial_rg_{tt}}{\partial_rg_{\phi\phi}}}=\sqrt{\frac{M(1+\alpha)}{r^3}\left(1-\frac{M\alpha}{r}\right)}\ ,
\end{align}
while in the Schwarzschild space it takes a quite simple form, $\Omega=\sqrt{M/r^3}$. Consequently, the orbital velocity measured by a local observer reads
\begin{align}\label{V}
v=\sqrt{\frac{g_{\phi\phi}}{g_{tt}}\frac{\partial_rg_{tt}}{\partial_rg_{\phi\phi}}}=\sqrt{\frac{M(1+\alpha)}{rf}\left(1-\frac{M\alpha}{r}\right)}\ .
\end{align}
One can easily check that test particle orbits around the Schwarzschild black hole at the ISCO position with a half of the speed of light, i.e. $v_{\rm ISCO}=1/2$, while the maximal orbital velocity of particle at the ISCO position around the Schwarzschild-MOG black hole reads 
\begin{align}
\lim_{\alpha\to\infty} v_{\rm ISCO}=\frac{1}{\sqrt{3}}\simeq 0.577\ .    
\end{align}

\begin{figure}
\includegraphics[width=\hsize]{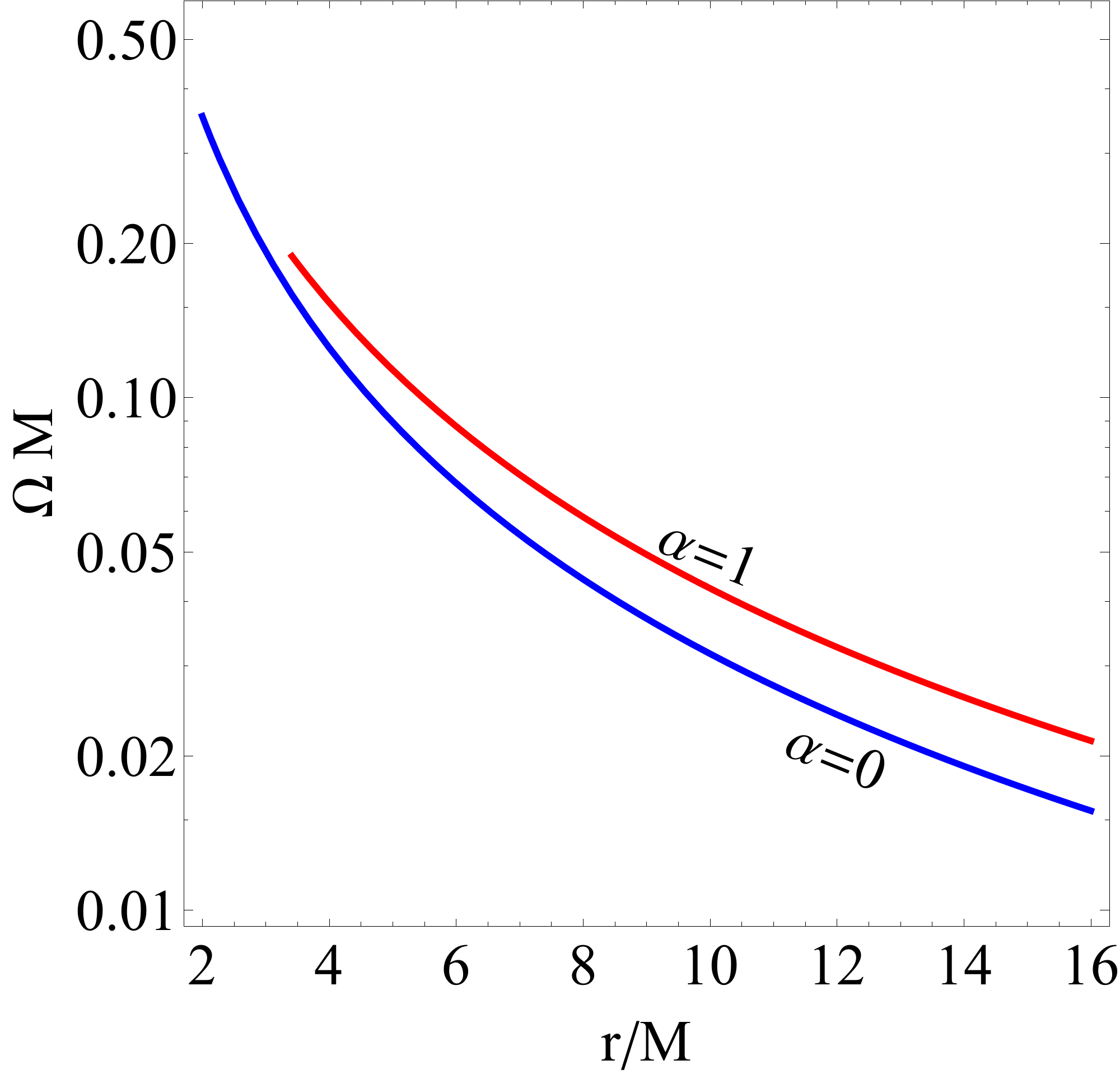}
\caption{Radial dependence of Keplerian angular frequency of test particle orbiting around the Schwarzschild-MOG black hole for the different values of $\alpha$ parameter.\label{Omega}}
\end{figure}

\begin{figure}
\includegraphics[width=\hsize]{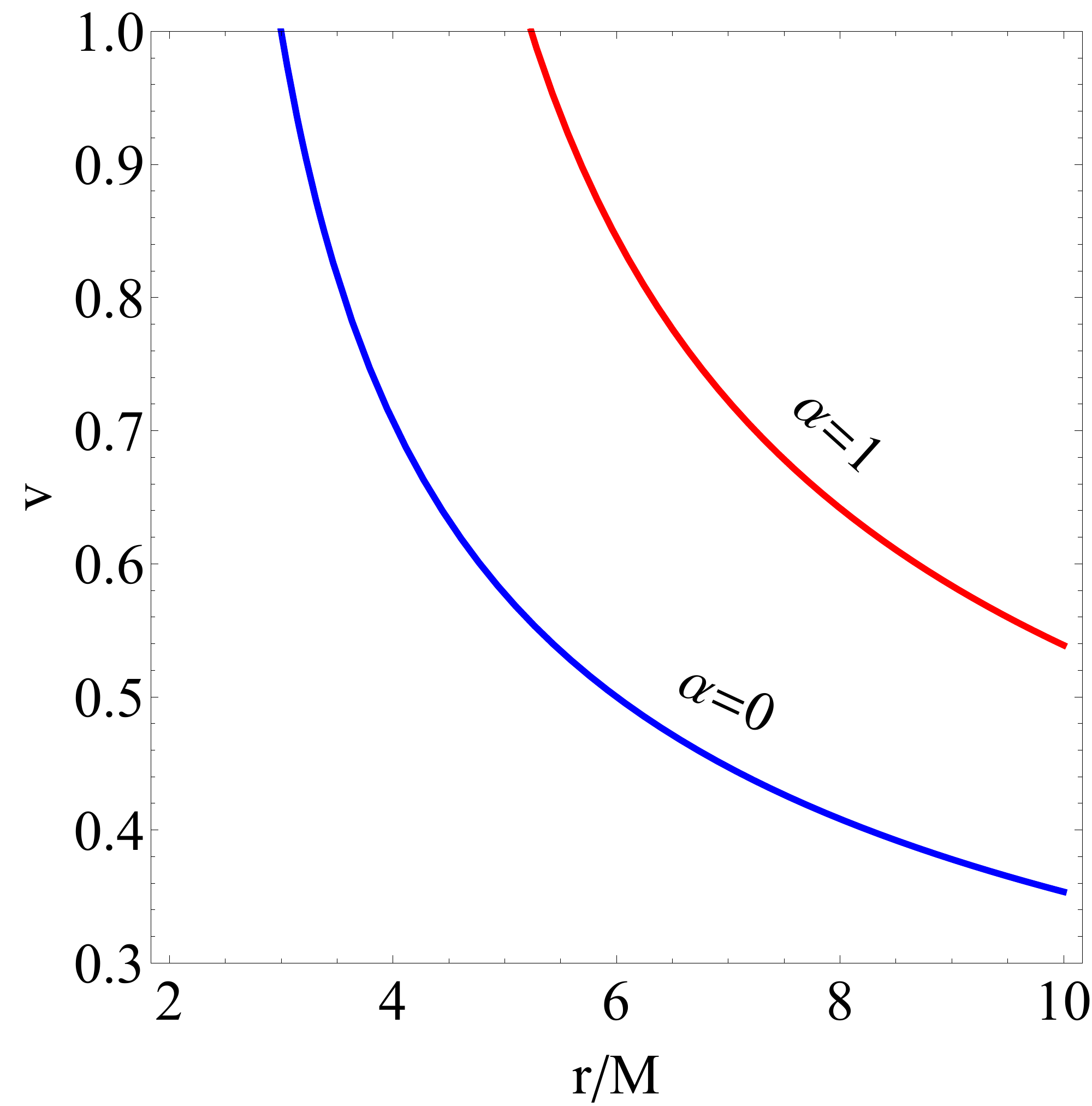}
\caption{Radial dependence of the orbital velocity of test particle orbiting around the Schwarzschild-MOG black hole for the different values of $\alpha$ parameter.\label{Vel}}
\end{figure}

\begin{figure}
\includegraphics[width=\hsize]{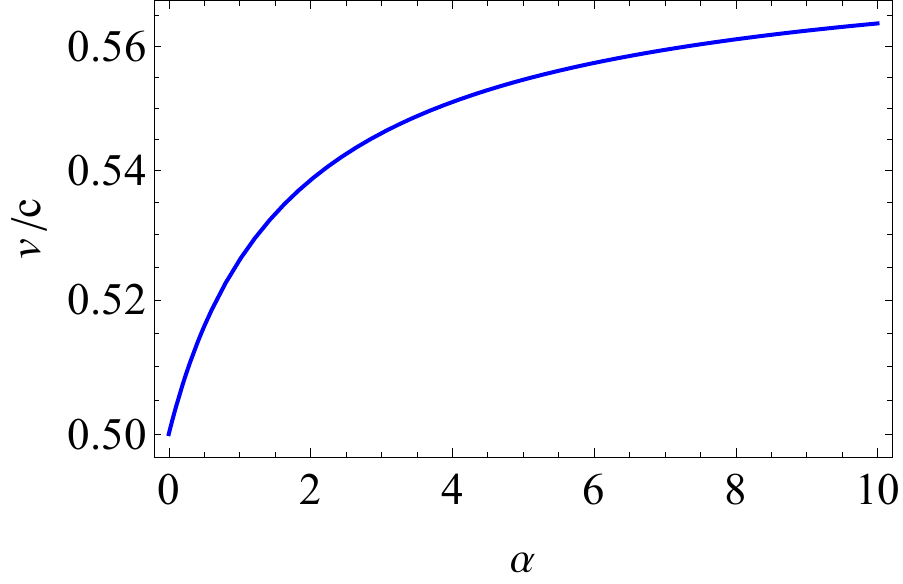}
\caption{The orbital velocity of test particle orbiting around the Schwarzschild black hole  at the ISCO position. The maximal orbital velocity of particle at the ISCO position reaches up to $v=1/\sqrt{3}$.\label{Vel2}}
\end{figure}

The radial dependence of the angular velocity $\Omega=\Omega(r)$ and linear velocity $v=v(r)$ is illustrated in Fig.~\ref{Omega} and Fig.~\ref{Vel}. In the case of Keplerian orbits, the angular velocity typically follows a $r^{-3/2}$ dependence. However, in the framework of MOG theory, this dependence is slightly modified due to the presence of the MOG parameter $\alpha$, as given by equation (\ref{W}). As a result, the angular frequency in MOG theory is larger than the Keplerian value.

Similarly, one can compare the linear velocity of test particles in different theories, namely general relativity and MOG. Fig.~\ref{Vel} demonstrates that the orbital velocity of test particles in MOG theory is greater than that in general relativity. To further comprehend the effects of MOG, one can examine the orbital velocity of test particles at the ISCO position. Fig.~\ref{Vel2} depicts the orbital velocity of particles at the ISCO position, revealing that for small values of the MOG parameter $\alpha$, test particles orbit around the black hole with a velocity of $v=1/2$. However, for the largest value of the MOG parameter, the orbital velocity of test particles at the ISCO position reaches up to $v=1/\sqrt{3}$, as shown in Fig.~\ref{Vel2}.

\section{Charged particle dynamics\label{Sec:ChargedParticleMotion}}

\subsection{Magnetic field configuration}

In the realistic astrophysical situation magnetic field's configuration in the vicinity of the gravitational compact object is very complex. However, for simplicity one can consider the analytical expression for the magnetic field, and one of the simple approach is performed by Wald~\cite{Wald74}. According to this approach the black hole is implanted in the asymptotically uniform magnetic field and the exact analytical expression for the vector potential in the Schwarzschild space is expressed as
\begin{align}\label{Uniform0}
A_\phi=\frac{1}{2}B r^2\sin^2\theta\ ,
\end{align}
where $B$ is the magnetic field strength. Note that the expression (\ref{Uniform0}) is fully satisfied the Maxwell equations in curved space given by
\begin{align}\label{ME}
\frac{1}{\sqrt{-g}}\partial_\alpha\left(\sqrt{-g}F^{\alpha\beta}\right)=0\ , \quad F_{\alpha\beta}=\partial_\alpha A_\beta-\partial_\beta A_\alpha\ .
\end{align}
Similarly, in the background of spacetime metric (\ref{metric}) the expression for the vector potential can be decomposed as
\begin{align}\label{Sol}
A_\phi=\frac{1}{2}B\psi(r)\sin^2\theta\ ,    
\end{align}
where $\psi(r)$ is radial function found as a solution of Maxwell equation (\ref{ME}). Inserting equation (\ref{Sol}) into (\ref{ME}), one can obtain   
\begin{align}\label{ME0}
r^2\left(f\psi'\right)'-2\psi=0\ .   
\end{align}
where a prime denotes the derivative with respect to radial coordinate. Finally, the exact analytical solution of Maxwell equation for the vector potential near the Schwarzschild-MOG black hole can be found as
\begin{align}
A_\phi=\frac{1}{2}B\left[r^2-\alpha(1+\alpha)M^2\right]\sin^2\theta\ , 
\end{align}
while non-zero components of the magnetic field measured by a proper observer read
\begin{align}
&B^{\hat r}=B\left(1+\frac{M^2\alpha(1+\alpha)}{r^2}\right)\cos\theta\ ,\\& B^{\hat\theta}=B\sqrt{f}\sin\theta\ .
\end{align}

\begin{figure*}
\includegraphics[width=\hsize]{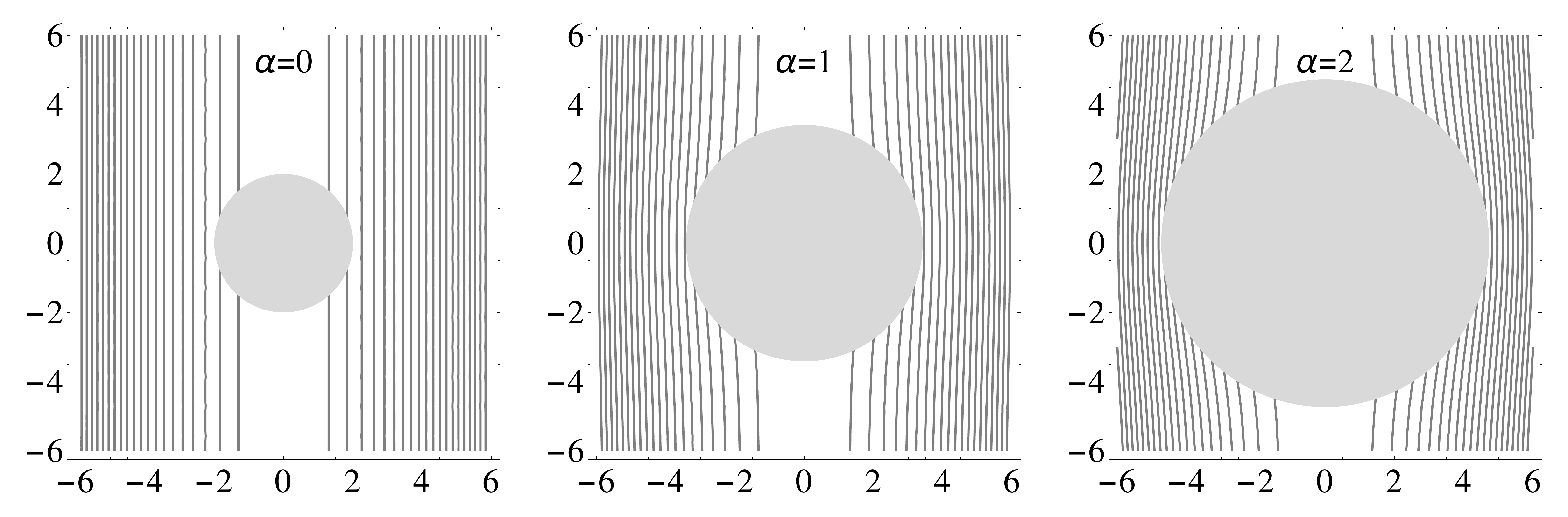}
\caption{The magnetic field lines in the vicinity of the Schwarzschild-MOG black hole for the different values of $\alpha$ parameter.\label{mag}}
\end{figure*}

To study the magnetic field configuration in the vicinity of the black hole, we can analyze the magnetic field lines represented by the equation $A_\phi=\text{const}$. Fig.~\ref{mag} provides a visualization of the magnetic field lines near the Schwarzschild-MOG black hole for various values of the $\alpha$ parameter. It is apparent that for larger values of $\alpha$, the magnetic field lines extend outward from the black hole in a more uniform manner.

\subsection{Charged particle motion}

Now one can study the charged particle motion around the Schwarzschild-MOG black hole immersed in external magnetic field. The Lagrangian for test particle of mass $m$ and charge $q$ can be written as
\begin{align} 
{\mathscr L}=\frac{1}{2}g_{\alpha\beta}{\dot x}^\alpha{\dot x}^\beta+\frac{q}{m}A_\alpha {\dot x}^\alpha\ , \qquad {\dot x}^\alpha=\frac{dx^\alpha}{ds}\ ,
\end{align}
and the constants of motion, namely, the specific energy, ${\cal E}$ and specific angular momentum, ${\cal L}$, of charged particle can be expressed as
\begin{align}
&p_t=\frac{\partial{\mathscr L}}{\partial\dot t}=g_{tt}{\dot t}=-{\cal E}\ ,\\  &p_\phi=\frac{\partial{\mathscr L}}{\partial\dot\phi}=g_{\phi\phi}{\dot\phi}+\frac{q}{m}A_\phi={\cal L}\ .    
\end{align}
Using the normalization of the 4-velocity $g_{\alpha\beta}{\dot x}^\alpha{\dot x}^\beta=-1$, one can obtain equation of the radial motion in the form:
\begin{align}
{\dot r}^2={\cal E}^2-V_{\rm eff}(r)\ ,
\end{align}
where the effective potential reads
\begin{align}\nonumber\label{Veff}
&V_{\rm eff}(\rho)=f(\rho)\left\{1+\frac{\left[{\cal L}-{\cal B}\left(\rho^2-\alpha(1+\alpha)\right)\right]^2}{\rho^2}\right\}\ , 
\end{align}
where $f(\rho)=1-2(1+\alpha)/\rho+\alpha(1+\alpha)/\rho^2$ with $\rho=r/M$ is the dimensionless radial coordinate and ${\cal B}$ is the magnetic parameter defined as
\begin{align}
{\cal B}=\frac{qBM}{2m}\ .    
\end{align}

Our analyses show that that in the presence of the external magnetic field the effective potential is divergent at the large distance:
\begin{align}
\lim_{r\to\infty} V_{\rm eff}(r)=\infty,     
\end{align}
In the previous section, it was mentioned that the magnetic field solution was obtained in the vicinity of the black hole, allowing us to safely use the effective potential in (\ref{Veff}) for the orbit of a charged particle in the vicinity of the black hole.

\begin{figure}
\includegraphics[width=\hsize]{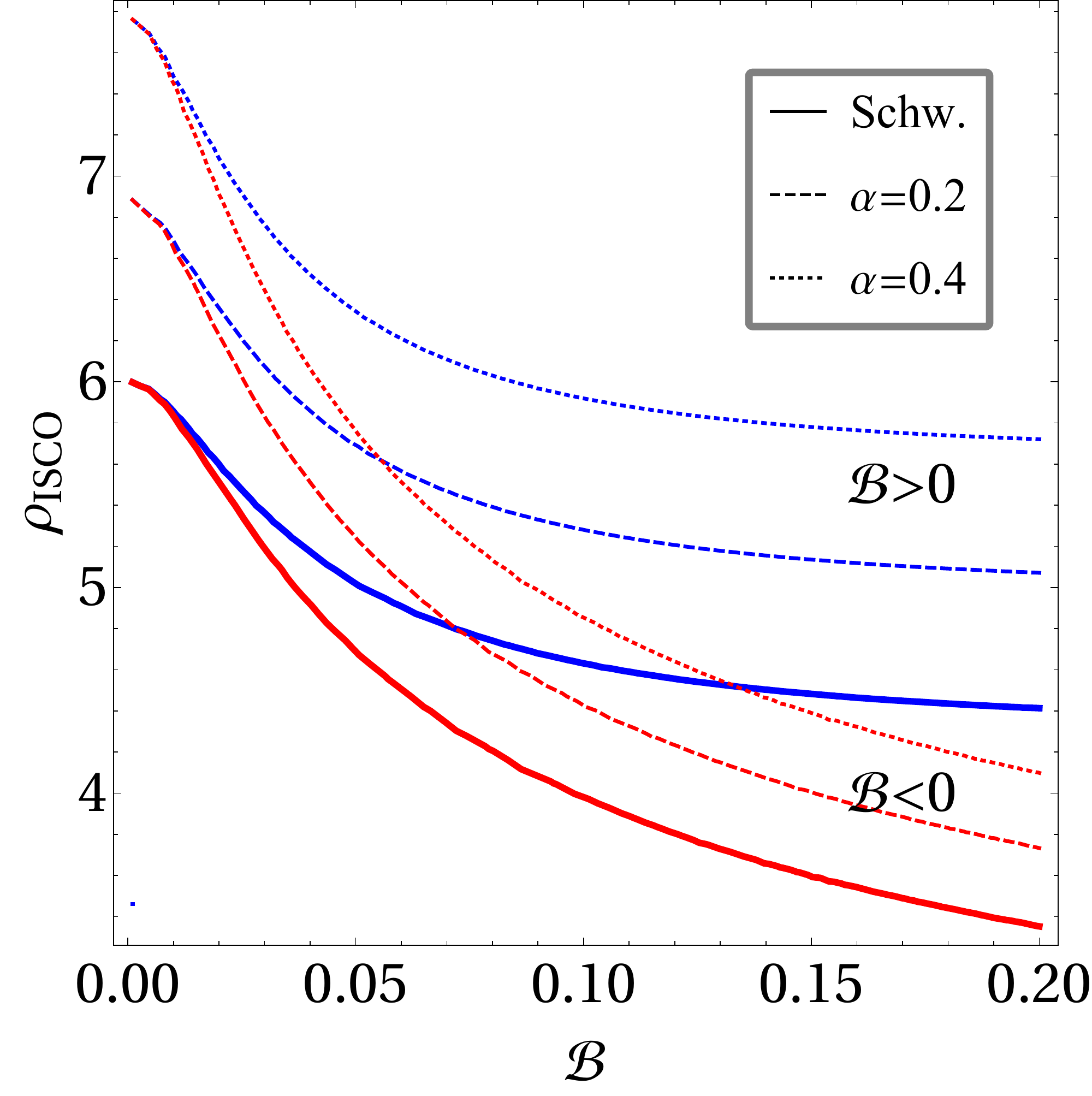}
\caption{Dependence of the ISCO position for charged particle orbiting around the Schwarzschild-MOG black hole from interaction parameter ${\cal B}$ for the different values of the MOG parameter $\alpha$. The solid blue line represent for the ISCO position for ${\cal B}>0$ while the solid red line represent for ${\cal B}>0$. The dashed and dotted blue and red lines are responsible the ISCO positions for charged particle for the different values of the MOG parameter $\alpha$. \label{ISCO}}
\end{figure}

By imposing the conditions ${\dot r}={\ddot r}=0$, the critical values of the specific energy and specific angular momentum for the charged particle can be determined. However, due to the complexity of the expressions, we will not present them here. Careful numerical analysis reveals that the innermost stable circular orbit (ISCO) position for the charged particle decreases in the presence of an external magnetic field. Fig.~\ref{ISCO} illustrates the dependence of the ISCO position for the charged particle on the magnetic parameter ${\cal B}$ for different values of the MOG parameter $\alpha$. The solid lines (top blue solid line for ${\cal B}>0$ and bottom red solid line for ${\cal B}<0$) represent the ISCO positions of the charged particle in the Schwarzschild spacetime (see, e.g., \cite{Kolos15}). We observe that the ISCO position of the test particle increases due to the MOG parameter, as shown in Fig.~\ref{Radii}. Similarly, the ISCO position for the charged particle increases with increasing values of the MOG parameter, as indicated by the dashed and dotted lines in Fig.~\ref{ISCO} for different values of the MOG parameter.

\section{Summary\label{Sec:Summary}}

In this research note, we have examined the geodesic motion in the Schwarzschild-MOG spacetime, focusing on the effects of the MOG parameter $\alpha$. Firstly, we analyzed photon motion around the black hole and derived analytical expressions for the photon sphere and impact parameter, which are crucial for understanding the black hole shadow. Our results revealed that both the photon sphere and the size of the black hole shadow increase due to the presence of MOG.

Next, we investigated the circular motion of massive particles in the Schwarzschild-MOG spacetime, specifically examining the innermost stable circular orbit (ISCO) and marginally bound orbit (MBO). We found that the characteristic radii of these orbits are larger compared to those in the pure Schwarzschild spacetime. Furthermore, we determined the maximum energy efficiency of test particles around the Schwarzschild-MOG black hole, which can reach up to approximately $8.14\%$, indicating a relatively large efficiency compared to the Schwarzschild case.

Considering the presence of an external asymptotically uniform magnetic field, proposed by Wald~\cite{Wald74}, we derived an exact analytical solution for the $A_\phi$ component of the vector potential of the electromagnetic field in the Schwarzschild-MOG background. The expressions for the radial and tangential components of the magnetic field were obtained. Our analysis revealed that the magnetic field exhibits a non-uniform behavior in the vicinity of the Schwarzschild-MOG black hole, with denser field lines compared to the pure Schwarzschild case.

Finally, we explored the motion of charged particles around the Schwarzschild-MOG black hole in the presence of the asymptotically uniform magnetic field. We specifically focused on circular motion and found that the ISCO position for charged particles is always smaller than that for neutral particles, emphasizing the influence of the external magnetic field on the orbits.

\section*{Acknowledgement}

This research is supported by Grants F-FA-2021-432, F-FA-2021-510, and MRB-2021-527 of the Uzbekistan Ministry for Innovative Development and by the Abdus Salam International Centre for Theoretical Physics under the Grant No. OEA-NT-01. AA thanks the PIFI fund of Chinese Academy of Sciences for the support.  

\bibliography{Ref,Shadow,reference}

%merlin.mbs apsrev4-1.bst 2010-07-25 4.21a (PWD, AO, DPC) hacked
%Control: key (0)
%Control: author (8) initials jnrlst
%Control: editor formatted (1) identically to author
%Control: production of article title (-1) disabled
%Control: page (0) single
%Control: year (1) truncated
%Control: production of eprint (0) enabled
\begin{thebibliography}{81}%
\makeatletter
\providecommand \@ifxundefined [1]{%
 \@ifx{#1\undefined}
}%
\providecommand \@ifnum [1]{%
 \ifnum #1\expandafter \@firstoftwo
 \else \expandafter \@secondoftwo
 \fi
}%
\providecommand \@ifx [1]{%
 \ifx #1\expandafter \@firstoftwo
 \else \expandafter \@secondoftwo
 \fi
}%
\providecommand \natexlab [1]{#1}%
\providecommand \enquote  [1]{``#1''}%
\providecommand \bibnamefont  [1]{#1}%
\providecommand \bibfnamefont [1]{#1}%
\providecommand \citenamefont [1]{#1}%
\providecommand \href@noop [0]{\@secondoftwo}%
\providecommand \href [0]{\begingroup \@sanitize@url \@href}%
\providecommand \@href[1]{\@@startlink{#1}\@@href}%
\providecommand \@@href[1]{\endgroup#1\@@endlink}%
\providecommand \@sanitize@url [0]{\catcode `\\12\catcode `\$12\catcode
  `\&12\catcode `\#12\catcode `\^12\catcode `\_12\catcode `\%12\relax}%
\providecommand \@@startlink[1]{}%
\providecommand \@@endlink[0]{}%
\providecommand \url  [0]{\begingroup\@sanitize@url \@url }%
\providecommand \@url [1]{\endgroup\@href {#1}{\urlprefix }}%
\providecommand \urlprefix  [0]{URL }%
\providecommand \Eprint [0]{\href }%
\providecommand \doibase [0]{http://dx.doi.org/}%
\providecommand \selectlanguage [0]{\@gobble}%
\providecommand \bibinfo  [0]{\@secondoftwo}%
\providecommand \bibfield  [0]{\@secondoftwo}%
\providecommand \translation [1]{[#1]}%
\providecommand \BibitemOpen [0]{}%
\providecommand \bibitemStop [0]{}%
\providecommand \bibitemNoStop [0]{.\EOS\space}%
\providecommand \EOS [0]{\spacefactor3000\relax}%
\providecommand \BibitemShut  [1]{\csname bibitem#1\endcsname}%
\let\auto@bib@innerbib\@empty
%</preamble>
\bibitem [{\citenamefont {{Bertone}}\ and\ \citenamefont
  {{Tait}}(2018)}]{Bertone2018Nature}%
  \BibitemOpen
  \bibfield  {author} {\bibinfo {author} {\bibfnamefont {G.}~\bibnamefont
  {{Bertone}}}\ and\ \bibinfo {author} {\bibfnamefont {T.~M.~P.}\ \bibnamefont
  {{Tait}}},\ }\href {\doibase 10.1038/s41586-018-0542-z} {\bibfield  {journal}
  {\bibinfo  {journal} {Nature}\ }\textbf {\bibinfo {volume} {562}},\ \bibinfo
  {pages} {51} (\bibinfo {year} {2018})},\ \Eprint
  {http://arxiv.org/abs/1810.01668} {arXiv:1810.01668 [astro-ph.CO]}
  \BibitemShut {NoStop}%
\bibitem [{\citenamefont {{Bertone}}\ and\ \citenamefont
  {{Hooper}}(2018)}]{Bertone2018RMP}%
  \BibitemOpen
  \bibfield  {author} {\bibinfo {author} {\bibfnamefont {G.}~\bibnamefont
  {{Bertone}}}\ and\ \bibinfo {author} {\bibfnamefont {D.}~\bibnamefont
  {{Hooper}}},\ }\href {\doibase 10.1103/RevModPhys.90.045002} {\bibfield
  {journal} {\bibinfo  {journal} {Reviews of Modern Physics}\ }\textbf
  {\bibinfo {volume} {90}},\ \bibinfo {eid} {045002} (\bibinfo {year}
  {2018})},\ \Eprint {http://arxiv.org/abs/1605.04909} {arXiv:1605.04909
  [astro-ph.CO]} \BibitemShut {NoStop}%
\bibitem [{\citenamefont {{Moffat}}(2006)}]{Moffat2006CQG}%
  \BibitemOpen
  \bibfield  {author} {\bibinfo {author} {\bibfnamefont {J.~W.}\ \bibnamefont
  {{Moffat}}},\ }\href {\doibase 10.1088/0264-9381/23/23/009} {\bibfield
  {journal} {\bibinfo  {journal} {Classical and Quantum Gravity}\ }\textbf
  {\bibinfo {volume} {23}},\ \bibinfo {pages} {6767} (\bibinfo {year}
  {2006})},\ \Eprint {http://arxiv.org/abs/gr-qc/0605141} {arXiv:gr-qc/0605141
  [gr-qc]} \BibitemShut {NoStop}%
\bibitem [{\citenamefont {{Moffat}}\ and\ \citenamefont
  {{Rahvar}}(2013{\natexlab{a}})}]{Moffat13}%
  \BibitemOpen
  \bibfield  {author} {\bibinfo {author} {\bibfnamefont {J.~W.}\ \bibnamefont
  {{Moffat}}}\ and\ \bibinfo {author} {\bibfnamefont {S.}~\bibnamefont
  {{Rahvar}}},\ }\href {\doibase 10.1093/mnras/stt1670} {\bibfield  {journal}
  {\bibinfo  {journal} {Mon. Not. R. Astron. Soc}\ }\textbf {\bibinfo {volume}
  {436}},\ \bibinfo {pages} {1439} (\bibinfo {year} {2013}{\natexlab{a}})},\
  \Eprint {http://arxiv.org/abs/1306.6383} {arXiv:1306.6383 [astro-ph.GA]}
  \BibitemShut {NoStop}%
\bibitem [{\citenamefont {{Moffat}}(2015{\natexlab{a}})}]{Moffat15}%
  \BibitemOpen
  \bibfield  {author} {\bibinfo {author} {\bibfnamefont {J.~W.}\ \bibnamefont
  {{Moffat}}},\ }\href {\doibase 10.1140/epjc/s10052-015-3405-x} {\bibfield
  {journal} {\bibinfo  {journal} {European Physical Journal C}\ }\textbf
  {\bibinfo {volume} {75}},\ \bibinfo {pages} {175} (\bibinfo {year}
  {2015}{\natexlab{a}})},\ \Eprint {http://arxiv.org/abs/1412.5424}
  {arXiv:1412.5424 [gr-qc]} \BibitemShut {NoStop}%
\bibitem [{\citenamefont {{Moffat}}\ and\ \citenamefont
  {{Toth}}(2015)}]{Moffat15a}%
  \BibitemOpen
  \bibfield  {author} {\bibinfo {author} {\bibfnamefont {J.~W.}\ \bibnamefont
  {{Moffat}}}\ and\ \bibinfo {author} {\bibfnamefont {V.~T.}\ \bibnamefont
  {{Toth}}},\ }\href {\doibase 10.1103/PhysRevD.91.043004} {\bibfield
  {journal} {\bibinfo  {journal} {Phys. Rev. D}\ }\textbf {\bibinfo {volume}
  {91}},\ \bibinfo {eid} {043004} (\bibinfo {year} {2015})},\ \Eprint
  {http://arxiv.org/abs/1411.6701} {arXiv:1411.6701 [astro-ph.GA]} \BibitemShut
  {NoStop}%
\bibitem [{\citenamefont {{Moffat}}(2015{\natexlab{b}})}]{Moffat2015EPJC}%
  \BibitemOpen
  \bibfield  {author} {\bibinfo {author} {\bibfnamefont {J.~W.}\ \bibnamefont
  {{Moffat}}},\ }\href {\doibase 10.1140/epjc/s10052-015-3352-6} {\bibfield
  {journal} {\bibinfo  {journal} {European Physical Journal C}\ }\textbf
  {\bibinfo {volume} {75}},\ \bibinfo {eid} {130} (\bibinfo {year}
  {2015}{\natexlab{b}})},\ \Eprint {http://arxiv.org/abs/1502.01677}
  {arXiv:1502.01677 [gr-qc]} \BibitemShut {NoStop}%
\bibitem [{\citenamefont {{Moffat}}\ and\ \citenamefont
  {{Toth}}(2020)}]{Moffat2020PRD}%
  \BibitemOpen
  \bibfield  {author} {\bibinfo {author} {\bibfnamefont {J.~W.}\ \bibnamefont
  {{Moffat}}}\ and\ \bibinfo {author} {\bibfnamefont {V.~T.}\ \bibnamefont
  {{Toth}}},\ }\href {\doibase 10.1103/PhysRevD.101.024014} {\bibfield
  {journal} {\bibinfo  {journal} {Phys. Rev. D}\ }\textbf {\bibinfo {volume}
  {101}},\ \bibinfo {eid} {024014} (\bibinfo {year} {2020})},\ \Eprint
  {http://arxiv.org/abs/1904.04142} {arXiv:1904.04142 [gr-qc]} \BibitemShut
  {NoStop}%
\bibitem [{\citenamefont {{Manfredi}}\ \emph {et~al.}(2018)\citenamefont
  {{Manfredi}}, \citenamefont {{Mureika}},\ and\ \citenamefont
  {{Moffat}}}]{Manfredi2018PLB}%
  \BibitemOpen
  \bibfield  {author} {\bibinfo {author} {\bibfnamefont {L.}~\bibnamefont
  {{Manfredi}}}, \bibinfo {author} {\bibfnamefont {J.}~\bibnamefont
  {{Mureika}}}, \ and\ \bibinfo {author} {\bibfnamefont {J.}~\bibnamefont
  {{Moffat}}},\ }\href {\doibase 10.1016/j.physletb.2017.11.006} {\bibfield
  {journal} {\bibinfo  {journal} {Physics Letters B}\ }\textbf {\bibinfo
  {volume} {779}},\ \bibinfo {pages} {492} (\bibinfo {year} {2018})},\ \Eprint
  {http://arxiv.org/abs/1711.03199} {arXiv:1711.03199 [gr-qc]} \BibitemShut
  {NoStop}%
\bibitem [{\citenamefont {{Wondrak}}\ \emph {et~al.}(2018)\citenamefont
  {{Wondrak}}, \citenamefont {{Nicolini}},\ and\ \citenamefont
  {{Moffat}}}]{Wondrak18}%
  \BibitemOpen
  \bibfield  {author} {\bibinfo {author} {\bibfnamefont {M.~F.}\ \bibnamefont
  {{Wondrak}}}, \bibinfo {author} {\bibfnamefont {P.}~\bibnamefont
  {{Nicolini}}}, \ and\ \bibinfo {author} {\bibfnamefont {J.~W.}\ \bibnamefont
  {{Moffat}}},\ }\href {\doibase 10.1088/1475-7516/2018/12/021} {\bibfield
  {journal} {\bibinfo  {journal} {JCAP}\ }\textbf {\bibinfo {volume} {2018}},\
  \bibinfo {eid} {021} (\bibinfo {year} {2018})},\ \Eprint
  {http://arxiv.org/abs/1809.07509} {arXiv:1809.07509 [gr-qc]} \BibitemShut
  {NoStop}%
\bibitem [{\citenamefont {{Moffat}}\ and\ \citenamefont
  {{Toth}}(2009)}]{Moffat09}%
  \BibitemOpen
  \bibfield  {author} {\bibinfo {author} {\bibfnamefont {J.~W.}\ \bibnamefont
  {{Moffat}}}\ and\ \bibinfo {author} {\bibfnamefont {V.~T.}\ \bibnamefont
  {{Toth}}},\ }\href {\doibase 10.1111/j.1365-2966.2009.14876.x} {\bibfield
  {journal} {\bibinfo  {journal} {Mon. Not. R. Astron. Soc.}\ }\textbf
  {\bibinfo {volume} {397}},\ \bibinfo {pages} {1885} (\bibinfo {year}
  {2009})},\ \Eprint {http://arxiv.org/abs/0805.4774} {arXiv:0805.4774
  [astro-ph]} \BibitemShut {NoStop}%
\bibitem [{\citenamefont {{Rahvar}}\ and\ \citenamefont
  {{Moffat}}(2019)}]{Rahvar2019MNRAS}%
  \BibitemOpen
  \bibfield  {author} {\bibinfo {author} {\bibfnamefont {S.}~\bibnamefont
  {{Rahvar}}}\ and\ \bibinfo {author} {\bibfnamefont {J.~W.}\ \bibnamefont
  {{Moffat}}},\ }\href {\doibase 10.1093/mnras/sty3002} {\bibfield  {journal}
  {\bibinfo  {journal} {Mon.Not.R.Astron.Soc.}\ }\textbf {\bibinfo {volume}
  {482}},\ \bibinfo {pages} {4514} (\bibinfo {year} {2019})},\ \Eprint
  {http://arxiv.org/abs/1807.07424} {arXiv:1807.07424 [gr-qc]} \BibitemShut
  {NoStop}%
\bibitem [{\citenamefont {{Della Monica}}\ \emph
  {et~al.}(2022{\natexlab{a}})\citenamefont {{Della Monica}}, \citenamefont
  {{de Martino}},\ and\ \citenamefont {{de Laurentis}}}]{DellaMonica2022MNRAS}%
  \BibitemOpen
  \bibfield  {author} {\bibinfo {author} {\bibfnamefont {R.}~\bibnamefont
  {{Della Monica}}}, \bibinfo {author} {\bibfnamefont {I.}~\bibnamefont {{de
  Martino}}}, \ and\ \bibinfo {author} {\bibfnamefont {M.}~\bibnamefont {{de
  Laurentis}}},\ }\href {\doibase 10.1093/mnras/stab3727} {\bibfield  {journal}
  {\bibinfo  {journal} {Mon.Not.R.Astron.Soc.}\ }\textbf {\bibinfo {volume}
  {510}},\ \bibinfo {pages} {4757} (\bibinfo {year} {2022}{\natexlab{a}})},\
  \Eprint {http://arxiv.org/abs/2105.12687} {arXiv:2105.12687 [gr-qc]}
  \BibitemShut {NoStop}%
\bibitem [{\citenamefont {{Della Monica}}\ \emph
  {et~al.}(2022{\natexlab{b}})\citenamefont {{Della Monica}}, \citenamefont
  {{de Martino}},\ and\ \citenamefont {{de Laurentis}}}]{2022Univ}%
  \BibitemOpen
  \bibfield  {author} {\bibinfo {author} {\bibfnamefont {R.}~\bibnamefont
  {{Della Monica}}}, \bibinfo {author} {\bibfnamefont {I.}~\bibnamefont {{de
  Martino}}}, \ and\ \bibinfo {author} {\bibfnamefont {M.}~\bibnamefont {{de
  Laurentis}}},\ }\href {\doibase 10.3390/universe8020137} {\bibfield
  {journal} {\bibinfo  {journal} {Universe}\ }\textbf {\bibinfo {volume} {8}},\
  \bibinfo {pages} {137} (\bibinfo {year} {2022}{\natexlab{b}})},\ \Eprint
  {http://arxiv.org/abs/2206.12699} {arXiv:2206.12699 [gr-qc]} \BibitemShut
  {NoStop}%
\bibitem [{\citenamefont {{Turimov}}(2022)}]{Turimov2022MNRAS}%
  \BibitemOpen
  \bibfield  {author} {\bibinfo {author} {\bibfnamefont {B.~V.}\ \bibnamefont
  {{Turimov}}},\ }\href {\doibase 10.1093/mnras/stac2113} {\bibfield  {journal}
  {\bibinfo  {journal} {Mon.Not.R.Astron.Soc.}\ }\textbf {\bibinfo {volume}
  {516}},\ \bibinfo {pages} {434} (\bibinfo {year} {2022})}\BibitemShut
  {NoStop}%
\bibitem [{\citenamefont {{Haydarov}}\ \emph {et~al.}(2020)\citenamefont
  {{Haydarov}}, \citenamefont {{Rayimbaev}}, \citenamefont {{Abdujabbarov}},
  \citenamefont {{Palvanov}},\ and\ \citenamefont
  {{Begmatova}}}]{Haydarov2020EPJC}%
  \BibitemOpen
  \bibfield  {author} {\bibinfo {author} {\bibfnamefont {K.}~\bibnamefont
  {{Haydarov}}}, \bibinfo {author} {\bibfnamefont {J.}~\bibnamefont
  {{Rayimbaev}}}, \bibinfo {author} {\bibfnamefont {A.}~\bibnamefont
  {{Abdujabbarov}}}, \bibinfo {author} {\bibfnamefont {S.}~\bibnamefont
  {{Palvanov}}}, \ and\ \bibinfo {author} {\bibfnamefont {D.}~\bibnamefont
  {{Begmatova}}},\ }\href {\doibase 10.1140/epjc/s10052-020-7992-9} {\bibfield
  {journal} {\bibinfo  {journal} {European Physical Journal C}\ }\textbf
  {\bibinfo {volume} {80}},\ \bibinfo {eid} {399} (\bibinfo {year} {2020})},\
  \Eprint {http://arxiv.org/abs/2004.14868} {arXiv:2004.14868 [gr-qc]}
  \BibitemShut {NoStop}%
\bibitem [{\citenamefont {{Jawad}}\ \emph {et~al.}(2016)\citenamefont
  {{Jawad}}, \citenamefont {{Ali}}, \citenamefont {{Jamil}},\ and\
  \citenamefont {{Debnath}}}]{Jawad16}%
  \BibitemOpen
  \bibfield  {author} {\bibinfo {author} {\bibfnamefont {A.}~\bibnamefont
  {{Jawad}}}, \bibinfo {author} {\bibfnamefont {F.}~\bibnamefont {{Ali}}},
  \bibinfo {author} {\bibfnamefont {M.}~\bibnamefont {{Jamil}}}, \ and\
  \bibinfo {author} {\bibfnamefont {U.}~\bibnamefont {{Debnath}}},\ }\href
  {\doibase 10.1088/0253-6102/66/5/509} {\bibfield  {journal} {\bibinfo
  {journal} {Communications in Theoretical Physics}\ }\textbf {\bibinfo
  {volume} {66}},\ \bibinfo {pages} {509} (\bibinfo {year} {2016})},\ \Eprint
  {http://arxiv.org/abs/1610.07411} {arXiv:1610.07411 [gr-qc]} \BibitemShut
  {NoStop}%
\bibitem [{\citenamefont {{Hussain}}\ and\ \citenamefont
  {{Jamil}}(2015)}]{Hussain15}%
  \BibitemOpen
  \bibfield  {author} {\bibinfo {author} {\bibfnamefont {S.}~\bibnamefont
  {{Hussain}}}\ and\ \bibinfo {author} {\bibfnamefont {M.}~\bibnamefont
  {{Jamil}}},\ }\href {\doibase 10.1103/PhysRevD.92.043008} {\bibfield
  {journal} {\bibinfo  {journal} {Phys. Rev. D}\ }\textbf {\bibinfo {volume}
  {92}},\ \bibinfo {eid} {043008} (\bibinfo {year} {2015})},\ \Eprint
  {http://arxiv.org/abs/1508.02123} {arXiv:1508.02123 [gr-qc]} \BibitemShut
  {NoStop}%
\bibitem [{\citenamefont {{Babar}}\ \emph {et~al.}(2016)\citenamefont
  {{Babar}}, \citenamefont {{Jamil}},\ and\ \citenamefont {{Lim}}}]{Babar16}%
  \BibitemOpen
  \bibfield  {author} {\bibinfo {author} {\bibfnamefont {G.~Z.}\ \bibnamefont
  {{Babar}}}, \bibinfo {author} {\bibfnamefont {M.}~\bibnamefont {{Jamil}}}, \
  and\ \bibinfo {author} {\bibfnamefont {Y.-K.}\ \bibnamefont {{Lim}}},\ }\href
  {\doibase 10.1142/S0218271816500243} {\bibfield  {journal} {\bibinfo
  {journal} {International Journal of Modern Physics D}\ }\textbf {\bibinfo
  {volume} {25}},\ \bibinfo {eid} {1650024} (\bibinfo {year} {2016})},\ \Eprint
  {http://arxiv.org/abs/1504.00072} {arXiv:1504.00072 [gr-qc]} \BibitemShut
  {NoStop}%
\bibitem [{\citenamefont {{Ba{\~n}ados}}\ \emph {et~al.}(2009)\citenamefont
  {{Ba{\~n}ados}}, \citenamefont {{Silk}},\ and\ \citenamefont
  {{West}}}]{Banados09}%
  \BibitemOpen
  \bibfield  {author} {\bibinfo {author} {\bibfnamefont {M.}~\bibnamefont
  {{Ba{\~n}ados}}}, \bibinfo {author} {\bibfnamefont {J.}~\bibnamefont
  {{Silk}}}, \ and\ \bibinfo {author} {\bibfnamefont {S.~M.}\ \bibnamefont
  {{West}}},\ }\href {\doibase 10.1103/PhysRevLett.103.111102} {\bibfield
  {journal} {\bibinfo  {journal} {Physical Review Letters}\ }\textbf {\bibinfo
  {volume} {103}},\ \bibinfo {eid} {111102} (\bibinfo {year}
  {2009})}\BibitemShut {NoStop}%
\bibitem [{\citenamefont {{Majeed}}\ and\ \citenamefont
  {{Jamil}}(2017)}]{Majeed17}%
  \BibitemOpen
  \bibfield  {author} {\bibinfo {author} {\bibfnamefont {B.}~\bibnamefont
  {{Majeed}}}\ and\ \bibinfo {author} {\bibfnamefont {M.}~\bibnamefont
  {{Jamil}}},\ }\href {\doibase 10.1142/S0218271817410176} {\bibfield
  {journal} {\bibinfo  {journal} {International Journal of Modern Physics D}\
  }\textbf {\bibinfo {volume} {26}},\ \bibinfo {eid} {1741017} (\bibinfo {year}
  {2017})},\ \Eprint {http://arxiv.org/abs/1705.04167} {arXiv:1705.04167
  [gr-qc]} \BibitemShut {NoStop}%
\bibitem [{\citenamefont {{Zakria}}\ and\ \citenamefont
  {{Jamil}}(2015)}]{Zakria15}%
  \BibitemOpen
  \bibfield  {author} {\bibinfo {author} {\bibfnamefont {A.}~\bibnamefont
  {{Zakria}}}\ and\ \bibinfo {author} {\bibfnamefont {M.}~\bibnamefont
  {{Jamil}}},\ }\href {\doibase 10.1007/JHEP05(2015)147} {\bibfield  {journal}
  {\bibinfo  {journal} {Journal of High Energy Physics}\ }\textbf {\bibinfo
  {volume} {2015}},\ \bibinfo {eid} {147} (\bibinfo {year} {2015})},\ \Eprint
  {http://arxiv.org/abs/1501.06306} {arXiv:1501.06306 [gr-qc]} \BibitemShut
  {NoStop}%
\bibitem [{\citenamefont {{Brevik}}\ and\ \citenamefont
  {{Jamil}}(2019)}]{Brevik19}%
  \BibitemOpen
  \bibfield  {author} {\bibinfo {author} {\bibfnamefont {I.}~\bibnamefont
  {{Brevik}}}\ and\ \bibinfo {author} {\bibfnamefont {M.}~\bibnamefont
  {{Jamil}}},\ }\href {\doibase 10.1142/S0219887819500300} {\bibfield
  {journal} {\bibinfo  {journal} {International Journal of Geometric Methods in
  Modern Physics}\ }\textbf {\bibinfo {volume} {16}},\ \bibinfo {eid} {1950030}
  (\bibinfo {year} {2019})},\ \Eprint {http://arxiv.org/abs/1901.00002}
  {arXiv:1901.00002 [gr-qc]} \BibitemShut {NoStop}%
\bibitem [{\citenamefont {{Chen}}\ \emph {et~al.}(2016)\citenamefont {{Chen}},
  \citenamefont {{Wang}},\ and\ \citenamefont {{Jing}}}]{Chen16}%
  \BibitemOpen
  \bibfield  {author} {\bibinfo {author} {\bibfnamefont {S.}~\bibnamefont
  {{Chen}}}, \bibinfo {author} {\bibfnamefont {M.}~\bibnamefont {{Wang}}}, \
  and\ \bibinfo {author} {\bibfnamefont {J.}~\bibnamefont {{Jing}}},\ }\href
  {\doibase 10.1007/JHEP09(2016)082} {\bibfield  {journal} {\bibinfo  {journal}
  {Journal of High Energy Physics}\ }\textbf {\bibinfo {volume} {2016}},\
  \bibinfo {eid} {82} (\bibinfo {year} {2016})},\ \Eprint
  {http://arxiv.org/abs/1604.02785} {arXiv:1604.02785 [gr-qc]} \BibitemShut
  {NoStop}%
\bibitem [{\citenamefont {{Hashimoto}}\ and\ \citenamefont
  {{Tanahashi}}(2017)}]{Hashimoto17}%
  \BibitemOpen
  \bibfield  {author} {\bibinfo {author} {\bibfnamefont {K.}~\bibnamefont
  {{Hashimoto}}}\ and\ \bibinfo {author} {\bibfnamefont {N.}~\bibnamefont
  {{Tanahashi}}},\ }\href {\doibase 10.1103/PhysRevD.95.024007} {\bibfield
  {journal} {\bibinfo  {journal} {Phys. rev. D}\ }\textbf {\bibinfo {volume}
  {95}},\ \bibinfo {eid} {024007} (\bibinfo {year} {2017})},\ \Eprint
  {http://arxiv.org/abs/1610.06070} {arXiv:1610.06070 [hep-th]} \BibitemShut
  {NoStop}%
\bibitem [{\citenamefont {{Dalui}}\ \emph {et~al.}(2019)\citenamefont
  {{Dalui}}, \citenamefont {{Majhi}},\ and\ \citenamefont
  {{Mishra}}}]{Dalui19}%
  \BibitemOpen
  \bibfield  {author} {\bibinfo {author} {\bibfnamefont {S.}~\bibnamefont
  {{Dalui}}}, \bibinfo {author} {\bibfnamefont {B.~R.}\ \bibnamefont
  {{Majhi}}}, \ and\ \bibinfo {author} {\bibfnamefont {P.}~\bibnamefont
  {{Mishra}}},\ }\href {\doibase 10.1016/j.physletb.2018.11.050} {\bibfield
  {journal} {\bibinfo  {journal} {Physics Letters B}\ }\textbf {\bibinfo
  {volume} {788}},\ \bibinfo {pages} {486} (\bibinfo {year} {2019})},\ \Eprint
  {http://arxiv.org/abs/1803.06527} {arXiv:1803.06527 [gr-qc]} \BibitemShut
  {NoStop}%
\bibitem [{\citenamefont {{Han}}(2008)}]{Han08}%
  \BibitemOpen
  \bibfield  {author} {\bibinfo {author} {\bibfnamefont {W.}~\bibnamefont
  {{Han}}},\ }\href {\doibase 10.1007/s10714-007-0598-9} {\bibfield  {journal}
  {\bibinfo  {journal} {General Relativity and Gravitation}\ }\textbf {\bibinfo
  {volume} {40}},\ \bibinfo {pages} {1831} (\bibinfo {year} {2008})},\ \Eprint
  {http://arxiv.org/abs/1006.2229} {arXiv:1006.2229 [gr-qc]} \BibitemShut
  {NoStop}%
\bibitem [{\citenamefont {{de Moura}}\ and\ \citenamefont
  {{Letelier}}(2000)}]{Moura00}%
  \BibitemOpen
  \bibfield  {author} {\bibinfo {author} {\bibfnamefont {A.~P.~S.}\
  \bibnamefont {{de Moura}}}\ and\ \bibinfo {author} {\bibfnamefont {P.~S.}\
  \bibnamefont {{Letelier}}},\ }\href {\doibase 10.1103/PhysRevE.61.6506}
  {\bibfield  {journal} {\bibinfo  {journal} {Phys. Rev. E}\ }\textbf {\bibinfo
  {volume} {61}},\ \bibinfo {pages} {6506} (\bibinfo {year} {2000})},\ \Eprint
  {http://arxiv.org/abs/chao-dyn/9910035} {arXiv:chao-dyn/9910035 [nlin.CD]}
  \BibitemShut {NoStop}%
\bibitem [{\citenamefont {{Wald}}(1974)}]{Wald74}%
  \BibitemOpen
  \bibfield  {author} {\bibinfo {author} {\bibfnamefont {R.~M.}\ \bibnamefont
  {{Wald}}},\ }\href {\doibase 10.1103/PhysRevD.10.1680} {\bibfield  {journal}
  {\bibinfo  {journal} {Phys. Rev. D.}\ }\textbf {\bibinfo {volume} {10}},\
  \bibinfo {pages} {1680} (\bibinfo {year} {1974})}\BibitemShut {NoStop}%
\bibitem [{\citenamefont {{Aliev}}\ \emph {et~al.}(1986)\citenamefont
  {{Aliev}}, \citenamefont {{Galtsov}},\ and\ \citenamefont
  {{Petukhov}}}]{Aliev86}%
  \BibitemOpen
  \bibfield  {author} {\bibinfo {author} {\bibfnamefont {A.~N.}\ \bibnamefont
  {{Aliev}}}, \bibinfo {author} {\bibfnamefont {D.~V.}\ \bibnamefont
  {{Galtsov}}}, \ and\ \bibinfo {author} {\bibfnamefont {V.~I.}\ \bibnamefont
  {{Petukhov}}},\ }\href {\doibase 10.1007/BF00649756} {\bibfield  {journal}
  {\bibinfo  {journal} {Astrophys. Space Sci.}\ }\textbf {\bibinfo {volume}
  {124}},\ \bibinfo {pages} {137} (\bibinfo {year} {1986})}\BibitemShut
  {NoStop}%
\bibitem [{\citenamefont {{Aliev}}\ and\ \citenamefont
  {{Gal'tsov}}(1989)}]{Aliev89}%
  \BibitemOpen
  \bibfield  {author} {\bibinfo {author} {\bibfnamefont {A.~N.}\ \bibnamefont
  {{Aliev}}}\ and\ \bibinfo {author} {\bibfnamefont {D.~V.}\ \bibnamefont
  {{Gal'tsov}}},\ }\href {\doibase 10.1070/PU1989v032n01ABEH002677} {\bibfield
  {journal} {\bibinfo  {journal} {Soviet Physics Uspekhi}\ }\textbf {\bibinfo
  {volume} {32}},\ \bibinfo {pages} {75} (\bibinfo {year} {1989})}\BibitemShut
  {NoStop}%
\bibitem [{\citenamefont {{Aliev}}\ and\ \citenamefont
  {{{\"O}zdemir}}(2002)}]{Aliev02}%
  \BibitemOpen
  \bibfield  {author} {\bibinfo {author} {\bibfnamefont {A.~N.}\ \bibnamefont
  {{Aliev}}}\ and\ \bibinfo {author} {\bibfnamefont {N.}~\bibnamefont
  {{{\"O}zdemir}}},\ }\href {\doibase 10.1046/j.1365-8711.2002.05727.x}
  {\bibfield  {journal} {\bibinfo  {journal} {Mon. Not. R. Astron. Soc.}\
  }\textbf {\bibinfo {volume} {336}},\ \bibinfo {pages} {241} (\bibinfo {year}
  {2002})},\ \Eprint {http://arxiv.org/abs/gr-qc/0208025} {gr-qc/0208025}
  \BibitemShut {NoStop}%
\bibitem [{\citenamefont {{Frolov}}\ and\ \citenamefont
  {{Krtou{\v{s}}}}(2011)}]{Frolov11}%
  \BibitemOpen
  \bibfield  {author} {\bibinfo {author} {\bibfnamefont {V.~P.}\ \bibnamefont
  {{Frolov}}}\ and\ \bibinfo {author} {\bibfnamefont {P.}~\bibnamefont
  {{Krtou{\v{s}}}}},\ }\href {\doibase 10.1103/PhysRevD.83.024016} {\bibfield
  {journal} {\bibinfo  {journal} {Phys. Rev. D}\ }\textbf {\bibinfo {volume}
  {83}},\ \bibinfo {eid} {024016} (\bibinfo {year} {2011})},\ \Eprint
  {http://arxiv.org/abs/1010.2266} {arXiv:1010.2266 [hep-th]} \BibitemShut
  {NoStop}%
\bibitem [{\citenamefont {{Frolov}}(2012)}]{Frolov12}%
  \BibitemOpen
  \bibfield  {author} {\bibinfo {author} {\bibfnamefont {V.~P.}\ \bibnamefont
  {{Frolov}}},\ }\href {\doibase 10.1103/PhysRevD.85.024020} {\bibfield
  {journal} {\bibinfo  {journal} {Phys. Rev. D.}\ }\textbf {\bibinfo {volume}
  {85}},\ \bibinfo {eid} {024020} (\bibinfo {year} {2012})},\ \Eprint
  {http://arxiv.org/abs/1110.6274} {arXiv:1110.6274 [gr-qc]} \BibitemShut
  {NoStop}%
\bibitem [{\citenamefont {Benavides-Gallego}\ \emph {et~al.}(2019)\citenamefont
  {Benavides-Gallego}, \citenamefont {Abdujabbarov}, \citenamefont
  {Malafarina}, \citenamefont {Ahmedov},\ and\ \citenamefont
  {Bambi}}]{Benavides-Gallego18}%
  \BibitemOpen
  \bibfield  {author} {\bibinfo {author} {\bibfnamefont {C.~A.}\ \bibnamefont
  {Benavides-Gallego}}, \bibinfo {author} {\bibfnamefont {A.}~\bibnamefont
  {Abdujabbarov}}, \bibinfo {author} {\bibfnamefont {D.}~\bibnamefont
  {Malafarina}}, \bibinfo {author} {\bibfnamefont {B.}~\bibnamefont {Ahmedov}},
  \ and\ \bibinfo {author} {\bibfnamefont {C.}~\bibnamefont {Bambi}},\ }\href
  {\doibase 10.1103/PhysRevD.99.044012} {\bibfield  {journal} {\bibinfo
  {journal} {Phys. Rev. D}\ }\textbf {\bibinfo {volume} {99}},\ \bibinfo
  {pages} {044012} (\bibinfo {year} {2019})},\ \Eprint
  {http://arxiv.org/abs/1812.04846} {arXiv:1812.04846 [gr-qc]} \BibitemShut
  {NoStop}%
%%CITATION = ARXIV:1812.04846;%%
\bibitem [{\citenamefont {{Shaymatov}}\ \emph {et~al.}(2018)\citenamefont
  {{Shaymatov}}, \citenamefont {{Ahmedov}}, \citenamefont {{Stuchl{\'{\i}}k}},\
  and\ \citenamefont {{Abdujabbarov}}}]{Shaymatov18}%
  \BibitemOpen
  \bibfield  {author} {\bibinfo {author} {\bibfnamefont {S.}~\bibnamefont
  {{Shaymatov}}}, \bibinfo {author} {\bibfnamefont {B.}~\bibnamefont
  {{Ahmedov}}}, \bibinfo {author} {\bibfnamefont {Z.}~\bibnamefont
  {{Stuchl{\'{\i}}k}}}, \ and\ \bibinfo {author} {\bibfnamefont
  {A.}~\bibnamefont {{Abdujabbarov}}},\ }\href {\doibase
  10.1142/S0218271818500888} {\bibfield  {journal} {\bibinfo  {journal}
  {International Journal of Modern Physics D}\ }\textbf {\bibinfo {volume}
  {27}},\ \bibinfo {eid} {1850088} (\bibinfo {year} {2018})}\BibitemShut
  {NoStop}%
\bibitem [{\citenamefont {{Stuchl{\'{\i}}k}}\ \emph {et~al.}(2014)\citenamefont
  {{Stuchl{\'{\i}}k}}, \citenamefont {{Schee}},\ and\ \citenamefont
  {{Abdujabbarov}}}]{Stuchlik14a}%
  \BibitemOpen
  \bibfield  {author} {\bibinfo {author} {\bibfnamefont {Z.}~\bibnamefont
  {{Stuchl{\'{\i}}k}}}, \bibinfo {author} {\bibfnamefont {J.}~\bibnamefont
  {{Schee}}}, \ and\ \bibinfo {author} {\bibfnamefont {A.}~\bibnamefont
  {{Abdujabbarov}}},\ }\href {\doibase 10.1103/PhysRevD.89.104048} {\bibfield
  {journal} {\bibinfo  {journal} {Phys. Rev. D}\ }\textbf {\bibinfo {volume}
  {89}},\ \bibinfo {eid} {104048} (\bibinfo {year} {2014})}\BibitemShut
  {NoStop}%
\bibitem [{\citenamefont {{Abdujabbarov}}\ and\ \citenamefont
  {{Ahmedov}}(2010)}]{Abdujabbarov10}%
  \BibitemOpen
  \bibfield  {author} {\bibinfo {author} {\bibfnamefont {A.}~\bibnamefont
  {{Abdujabbarov}}}\ and\ \bibinfo {author} {\bibfnamefont {B.}~\bibnamefont
  {{Ahmedov}}},\ }\href {\doibase 10.1103/PhysRevD.81.044022} {\bibfield
  {journal} {\bibinfo  {journal} {Phys. Rev. D}\ }\textbf {\bibinfo {volume}
  {81}},\ \bibinfo {eid} {044022} (\bibinfo {year} {2010})},\ \Eprint
  {http://arxiv.org/abs/0905.2730} {arXiv:0905.2730 [gr-qc]} \BibitemShut
  {NoStop}%
\bibitem [{\citenamefont {{Abdujabbarov}}\ \emph
  {et~al.}(2011{\natexlab{a}})\citenamefont {{Abdujabbarov}}, \citenamefont
  {{Ahmedov}},\ and\ \citenamefont {{Hakimov}}}]{Abdujabbarov11a}%
  \BibitemOpen
  \bibfield  {author} {\bibinfo {author} {\bibfnamefont {A.}~\bibnamefont
  {{Abdujabbarov}}}, \bibinfo {author} {\bibfnamefont {B.}~\bibnamefont
  {{Ahmedov}}}, \ and\ \bibinfo {author} {\bibfnamefont {A.}~\bibnamefont
  {{Hakimov}}},\ }\href {\doibase 10.1103/PhysRevD.83.044053} {\bibfield
  {journal} {\bibinfo  {journal} {Phys.Rev. D}\ }\textbf {\bibinfo {volume}
  {83}},\ \bibinfo {eid} {044053} (\bibinfo {year} {2011}{\natexlab{a}})},\
  \Eprint {http://arxiv.org/abs/1101.4741} {arXiv:1101.4741 [gr-qc]}
  \BibitemShut {NoStop}%
\bibitem [{\citenamefont {{Abdujabbarov}}\ \emph
  {et~al.}(2011{\natexlab{b}})\citenamefont {{Abdujabbarov}}, \citenamefont
  {{Ahmedov}}, \citenamefont {{Shaymatov}},\ and\ \citenamefont
  {{Rakhmatov}}}]{Abdujabbarov11}%
  \BibitemOpen
  \bibfield  {author} {\bibinfo {author} {\bibfnamefont {A.~A.}\ \bibnamefont
  {{Abdujabbarov}}}, \bibinfo {author} {\bibfnamefont {B.~J.}\ \bibnamefont
  {{Ahmedov}}}, \bibinfo {author} {\bibfnamefont {S.~R.}\ \bibnamefont
  {{Shaymatov}}}, \ and\ \bibinfo {author} {\bibfnamefont {A.~S.}\ \bibnamefont
  {{Rakhmatov}}},\ }\href {\doibase 10.1007/s10509-011-0740-8} {\bibfield
  {journal} {\bibinfo  {journal} {Astrophys Space Sci}\ }\textbf {\bibinfo
  {volume} {334}},\ \bibinfo {pages} {237} (\bibinfo {year}
  {2011}{\natexlab{b}})},\ \Eprint {http://arxiv.org/abs/1105.1910}
  {arXiv:1105.1910 [astro-ph.SR]} \BibitemShut {NoStop}%
\bibitem [{\citenamefont {{Abdujabbarov}}\ \emph {et~al.}(2008)\citenamefont
  {{Abdujabbarov}}, \citenamefont {{Ahmedov}},\ and\ \citenamefont
  {{Kagramanova}}}]{Abdujabbarov08}%
  \BibitemOpen
  \bibfield  {author} {\bibinfo {author} {\bibfnamefont {A.~A.}\ \bibnamefont
  {{Abdujabbarov}}}, \bibinfo {author} {\bibfnamefont {B.~J.}\ \bibnamefont
  {{Ahmedov}}}, \ and\ \bibinfo {author} {\bibfnamefont {V.~G.}\ \bibnamefont
  {{Kagramanova}}},\ }\href {\doibase 10.1007/s10714-008-0635-3} {\bibfield
  {journal} {\bibinfo  {journal} {General Relativity and Gravitation}\ }\textbf
  {\bibinfo {volume} {40}},\ \bibinfo {pages} {2515} (\bibinfo {year}
  {2008})},\ \Eprint {http://arxiv.org/abs/0802.4349} {arXiv:0802.4349 [gr-qc]}
  \BibitemShut {NoStop}%
\bibitem [{\citenamefont {{Karas}}\ \emph {et~al.}(2012)\citenamefont
  {{Karas}}, \citenamefont {{Kovar}}, \citenamefont {{Kopacek}}, \citenamefont
  {{Kojima}}, \citenamefont {{Slany}},\ and\ \citenamefont
  {{Stuchlik}}}]{Karas12a}%
  \BibitemOpen
  \bibfield  {author} {\bibinfo {author} {\bibfnamefont {V.}~\bibnamefont
  {{Karas}}}, \bibinfo {author} {\bibfnamefont {J.}~\bibnamefont {{Kovar}}},
  \bibinfo {author} {\bibfnamefont {O.}~\bibnamefont {{Kopacek}}}, \bibinfo
  {author} {\bibfnamefont {Y.}~\bibnamefont {{Kojima}}}, \bibinfo {author}
  {\bibfnamefont {P.}~\bibnamefont {{Slany}}}, \ and\ \bibinfo {author}
  {\bibfnamefont {Z.}~\bibnamefont {{Stuchlik}}},\ }in\ \href@noop {} {\emph
  {\bibinfo {booktitle} {American Astronomical Society Meeting Abstracts
  \#220}}},\ \bibinfo {series} {American Astronomical Society Meeting
  Abstracts}, Vol.\ \bibinfo {volume} {220}\ (\bibinfo {year} {2012})\ p.\
  \bibinfo {pages} {430.07}\BibitemShut {NoStop}%
\bibitem [{\citenamefont {{Stuchl{\'{\i}}k}}\ and\ \citenamefont {{Kolo{\v
  s}}}(2016)}]{Stuchlik16}%
  \BibitemOpen
  \bibfield  {author} {\bibinfo {author} {\bibfnamefont {Z.}~\bibnamefont
  {{Stuchl{\'{\i}}k}}}\ and\ \bibinfo {author} {\bibfnamefont {M.}~\bibnamefont
  {{Kolo{\v s}}}},\ }\href {\doibase 10.1140/epjc/s10052-015-3862-2} {\bibfield
   {journal} {\bibinfo  {journal} {European Physical Journal C}\ }\textbf
  {\bibinfo {volume} {76}},\ \bibinfo {eid} {32} (\bibinfo {year} {2016})},\
  \Eprint {http://arxiv.org/abs/1511.02936} {arXiv:1511.02936 [gr-qc]}
  \BibitemShut {NoStop}%
\bibitem [{\citenamefont {{Kov{\'a}{\v r}}}\ \emph {et~al.}(2010)\citenamefont
  {{Kov{\'a}{\v r}}}, \citenamefont {{Kop{\'a}{\v c}ek}}, \citenamefont
  {{Karas}},\ and\ \citenamefont {{Stuchl{\'{\i}}k}}}]{Kovar10}%
  \BibitemOpen
  \bibfield  {author} {\bibinfo {author} {\bibfnamefont {J.}~\bibnamefont
  {{Kov{\'a}{\v r}}}}, \bibinfo {author} {\bibfnamefont {O.}~\bibnamefont
  {{Kop{\'a}{\v c}ek}}}, \bibinfo {author} {\bibfnamefont {V.}~\bibnamefont
  {{Karas}}}, \ and\ \bibinfo {author} {\bibfnamefont {Z.}~\bibnamefont
  {{Stuchl{\'{\i}}k}}},\ }\href {\doibase 10.1088/0264-9381/27/13/135006}
  {\bibfield  {journal} {\bibinfo  {journal} {Classical and Quantum Gravity}\
  }\textbf {\bibinfo {volume} {27}},\ \bibinfo {eid} {135006} (\bibinfo {year}
  {2010})},\ \Eprint {http://arxiv.org/abs/1005.3270} {arXiv:1005.3270
  [astro-ph.HE]} \BibitemShut {NoStop}%
\bibitem [{\citenamefont {{Kov{\'a}{\v r}}}\ \emph {et~al.}(2014)\citenamefont
  {{Kov{\'a}{\v r}}}, \citenamefont {{Slan{\'y}}}, \citenamefont
  {{Cremaschini}}, \citenamefont {{Stuchl{\'{\i}}k}}, \citenamefont {{Karas}},\
  and\ \citenamefont {{Trova}}}]{Kovar14}%
  \BibitemOpen
  \bibfield  {author} {\bibinfo {author} {\bibfnamefont {J.}~\bibnamefont
  {{Kov{\'a}{\v r}}}}, \bibinfo {author} {\bibfnamefont {P.}~\bibnamefont
  {{Slan{\'y}}}}, \bibinfo {author} {\bibfnamefont {C.}~\bibnamefont
  {{Cremaschini}}}, \bibinfo {author} {\bibfnamefont {Z.}~\bibnamefont
  {{Stuchl{\'{\i}}k}}}, \bibinfo {author} {\bibfnamefont {V.}~\bibnamefont
  {{Karas}}}, \ and\ \bibinfo {author} {\bibfnamefont {A.}~\bibnamefont
  {{Trova}}},\ }\href {\doibase 10.1103/PhysRevD.90.044029} {\bibfield
  {journal} {\bibinfo  {journal} {Phys. Rev. D}\ }\textbf {\bibinfo {volume}
  {90}},\ \bibinfo {eid} {044029} (\bibinfo {year} {2014})},\ \Eprint
  {http://arxiv.org/abs/1409.0418} {arXiv:1409.0418 [gr-qc]} \BibitemShut
  {NoStop}%
\bibitem [{\citenamefont {{Kolo{\v s}}}\ \emph {et~al.}(2017)\citenamefont
  {{Kolo{\v s}}}, \citenamefont {{Tursunov}},\ and\ \citenamefont
  {{Stuchl{\'{\i}}k}}}]{Kolos17}%
  \BibitemOpen
  \bibfield  {author} {\bibinfo {author} {\bibfnamefont {M.}~\bibnamefont
  {{Kolo{\v s}}}}, \bibinfo {author} {\bibfnamefont {A.}~\bibnamefont
  {{Tursunov}}}, \ and\ \bibinfo {author} {\bibfnamefont {Z.}~\bibnamefont
  {{Stuchl{\'{\i}}k}}},\ }\href@noop {} {\bibfield  {journal} {\bibinfo
  {journal} {Eur. Phys. J. C.}\ }\textbf {\bibinfo {volume} {77}},\ \bibinfo
  {pages} {860} (\bibinfo {year} {2017})},\ \Eprint
  {http://arxiv.org/abs/1707.02224} {arXiv:1707.02224 [astro-ph.HE]}
  \BibitemShut {NoStop}%
\bibitem [{\citenamefont {{Rayimbaev}}\ and\ \citenamefont
  {{Tadjimuratov}}(2020)}]{Pulat2020PhRvDMOG}%
  \BibitemOpen
  \bibfield  {author} {\bibinfo {author} {\bibfnamefont {J.}~\bibnamefont
  {{Rayimbaev}}}\ and\ \bibinfo {author} {\bibfnamefont {P.}~\bibnamefont
  {{Tadjimuratov}}},\ }\href {\doibase 10.1103/PhysRevD.102.024019} {\bibfield
  {journal} {\bibinfo  {journal} {Physical Review D}\ }\textbf {\bibinfo
  {volume} {102}},\ \bibinfo {eid} {024019} (\bibinfo {year}
  {2020})}\BibitemShut {NoStop}%
\bibitem [{\citenamefont {{Rayimbaev}}\ \emph
  {et~al.}(2019{\natexlab{a}})\citenamefont {{Rayimbaev}}, \citenamefont
  {{Turimov}},\ and\ \citenamefont {{Palvanov}}}]{Rayimbaev2019IJMPCS}%
  \BibitemOpen
  \bibfield  {author} {\bibinfo {author} {\bibfnamefont {J.}~\bibnamefont
  {{Rayimbaev}}}, \bibinfo {author} {\bibfnamefont {B.}~\bibnamefont
  {{Turimov}}}, \ and\ \bibinfo {author} {\bibfnamefont {S.}~\bibnamefont
  {{Palvanov}}},\ }in\ \href {\doibase 10.1142/S201019451960019X} {\emph
  {\bibinfo {booktitle} {International Journal of Modern Physics Conference
  Series}}},\ \bibinfo {series} {International Journal of Modern Physics
  Conference Series}, Vol.~\bibinfo {volume} {49}\ (\bibinfo {year} {2019})\
  pp.\ \bibinfo {pages} {1960019--209}\BibitemShut {NoStop}%
\bibitem [{\citenamefont {{Rayimbaev}}\ \emph
  {et~al.}(2020{\natexlab{a}})\citenamefont {{Rayimbaev}}, \citenamefont
  {{Turimov}}, \citenamefont {{Marcos}}, \citenamefont {{Palvanov}},\ and\
  \citenamefont {{Rakhmatov}}}]{Rayimbaev2020MPLA}%
  \BibitemOpen
  \bibfield  {author} {\bibinfo {author} {\bibfnamefont {J.}~\bibnamefont
  {{Rayimbaev}}}, \bibinfo {author} {\bibfnamefont {B.}~\bibnamefont
  {{Turimov}}}, \bibinfo {author} {\bibfnamefont {F.}~\bibnamefont {{Marcos}}},
  \bibinfo {author} {\bibfnamefont {S.}~\bibnamefont {{Palvanov}}}, \ and\
  \bibinfo {author} {\bibfnamefont {A.}~\bibnamefont {{Rakhmatov}}},\ }\href
  {\doibase 10.1142/S021773232050056X} {\bibfield  {journal} {\bibinfo
  {journal} {Modern Physics Letters A}\ }\textbf {\bibinfo {volume} {35}},\
  \bibinfo {eid} {2050056} (\bibinfo {year} {2020}{\natexlab{a}})}\BibitemShut
  {NoStop}%
\bibitem [{\citenamefont {{Rayimbaev}}\ \emph
  {et~al.}(2019{\natexlab{b}})\citenamefont {{Rayimbaev}}, \citenamefont
  {{Turimov}},\ and\ \citenamefont {{Ahmedov}}}]{Rayimbaev2019IJMPD}%
  \BibitemOpen
  \bibfield  {author} {\bibinfo {author} {\bibfnamefont {J.}~\bibnamefont
  {{Rayimbaev}}}, \bibinfo {author} {\bibfnamefont {B.}~\bibnamefont
  {{Turimov}}}, \ and\ \bibinfo {author} {\bibfnamefont {B.}~\bibnamefont
  {{Ahmedov}}},\ }\href {\doibase 10.1142/S0218271819501281} {\bibfield
  {journal} {\bibinfo  {journal} {International Journal of Modern Physics D}\
  }\textbf {\bibinfo {volume} {28}},\ \bibinfo {eid} {1950128-209} (\bibinfo
  {year} {2019}{\natexlab{b}})}\BibitemShut {NoStop}%
\bibitem [{\citenamefont {{de Felice}}\ and\ \citenamefont
  {{Sorge}}(2003)}]{deFelice}%
  \BibitemOpen
  \bibfield  {author} {\bibinfo {author} {\bibfnamefont {F.}~\bibnamefont {{de
  Felice}}}\ and\ \bibinfo {author} {\bibfnamefont {F.}~\bibnamefont
  {{Sorge}}},\ }\href@noop {} {\bibfield  {journal} {\bibinfo  {journal}
  {Classical and Quantum Gravity}\ }\textbf {\bibinfo {volume} {20}},\ \bibinfo
  {pages} {469} (\bibinfo {year} {2003})}\BibitemShut {NoStop}%
\bibitem [{\citenamefont {{de Felice}}\ \emph {et~al.}(2004)\citenamefont {{de
  Felice}}, \citenamefont {{Sorge}},\ and\ \citenamefont
  {{Zilio}}}]{deFelice2004}%
  \BibitemOpen
  \bibfield  {author} {\bibinfo {author} {\bibfnamefont {F.}~\bibnamefont {{de
  Felice}}}, \bibinfo {author} {\bibfnamefont {F.}~\bibnamefont {{Sorge}}}, \
  and\ \bibinfo {author} {\bibfnamefont {S.}~\bibnamefont {{Zilio}}},\ }\href
  {\doibase 10.1088/0264-9381/21/4/016} {\bibfield  {journal} {\bibinfo
  {journal} {Classical and Quantum Gravity}\ }\textbf {\bibinfo {volume}
  {21}},\ \bibinfo {pages} {961} (\bibinfo {year} {2004})}\BibitemShut
  {NoStop}%
\bibitem [{\citenamefont {{Rayimbaev}}(2016)}]{Rayimbaev16}%
  \BibitemOpen
  \bibfield  {author} {\bibinfo {author} {\bibfnamefont {J.~R.}\ \bibnamefont
  {{Rayimbaev}}},\ }\href {\doibase 10.1007/s10509-016-2879-9} {\bibfield
  {journal} {\bibinfo  {journal} {Astrophys Space Sc}\ }\textbf {\bibinfo
  {volume} {361}},\ \bibinfo {eid} {288} (\bibinfo {year} {2016})}\BibitemShut
  {NoStop}%
\bibitem [{\citenamefont {{Juraeva}}\ \emph {et~al.}(2021)\citenamefont
  {{Juraeva}}, \citenamefont {{Rayimbaev}}, \citenamefont {{Abdujabbarov}},
  \citenamefont {{Ahmedov}},\ and\ \citenamefont
  {{Palvanov}}}]{Juraeva2021EPJC}%
  \BibitemOpen
  \bibfield  {author} {\bibinfo {author} {\bibfnamefont {N.}~\bibnamefont
  {{Juraeva}}}, \bibinfo {author} {\bibfnamefont {J.}~\bibnamefont
  {{Rayimbaev}}}, \bibinfo {author} {\bibfnamefont {A.}~\bibnamefont
  {{Abdujabbarov}}}, \bibinfo {author} {\bibfnamefont {B.}~\bibnamefont
  {{Ahmedov}}}, \ and\ \bibinfo {author} {\bibfnamefont {S.}~\bibnamefont
  {{Palvanov}}},\ }\href {\doibase 10.1140/epjc/s10052-021-08876-5} {\bibfield
  {journal} {\bibinfo  {journal} {European Physical Journal C}\ }\textbf
  {\bibinfo {volume} {81}},\ \bibinfo {eid} {70} (\bibinfo {year}
  {2021})}\BibitemShut {NoStop}%
\bibitem [{\citenamefont {{Rayimbaev}}\ \emph {et~al.}(2021)\citenamefont
  {{Rayimbaev}}, \citenamefont {{Abdujabbarov}}, \citenamefont {{Jamil}},\ and\
  \citenamefont {{Han}}}]{Rayimbaev2021NuPhB}%
  \BibitemOpen
  \bibfield  {author} {\bibinfo {author} {\bibfnamefont {J.}~\bibnamefont
  {{Rayimbaev}}}, \bibinfo {author} {\bibfnamefont {A.}~\bibnamefont
  {{Abdujabbarov}}}, \bibinfo {author} {\bibfnamefont {M.}~\bibnamefont
  {{Jamil}}}, \ and\ \bibinfo {author} {\bibfnamefont {W.-B.}\ \bibnamefont
  {{Han}}},\ }\href {\doibase 10.1016/j.nuclphysb.2021.115364} {\bibfield
  {journal} {\bibinfo  {journal} {Nuclear Physics B}\ }\textbf {\bibinfo
  {volume} {966}},\ \bibinfo {eid} {115364} (\bibinfo {year} {2021})},\ \Eprint
  {http://arxiv.org/abs/2009.04898} {arXiv:2009.04898 [gr-qc]} \BibitemShut
  {NoStop}%
\bibitem [{\citenamefont {{Toshmatov}}\ \emph {et~al.}(2015)\citenamefont
  {{Toshmatov}}, \citenamefont {{Abdujabbarov}}, \citenamefont {{Ahmedov}},\
  and\ \citenamefont {{Stuchl{\'{\i}}k}}}]{Toshmatov15d}%
  \BibitemOpen
  \bibfield  {author} {\bibinfo {author} {\bibfnamefont {B.}~\bibnamefont
  {{Toshmatov}}}, \bibinfo {author} {\bibfnamefont {A.}~\bibnamefont
  {{Abdujabbarov}}}, \bibinfo {author} {\bibfnamefont {B.}~\bibnamefont
  {{Ahmedov}}}, \ and\ \bibinfo {author} {\bibfnamefont {Z.}~\bibnamefont
  {{Stuchl{\'{\i}}k}}},\ }\href {\doibase 10.1007/s10509-015-2533-y} {\bibfield
   {journal} {\bibinfo  {journal} {Astrophys Space Sci}\ }\textbf {\bibinfo
  {volume} {360}},\ \bibinfo {eid} {19} (\bibinfo {year} {2015})}\BibitemShut
  {NoStop}%
\bibitem [{\citenamefont {{Abdujabbarov}}\ \emph {et~al.}(2014)\citenamefont
  {{Abdujabbarov}}, \citenamefont {{Ahmedov}}, \citenamefont {{Rahimov}},\ and\
  \citenamefont {{Salikhbaev}}}]{Abdujabbarov14}%
  \BibitemOpen
  \bibfield  {author} {\bibinfo {author} {\bibfnamefont {A.}~\bibnamefont
  {{Abdujabbarov}}}, \bibinfo {author} {\bibfnamefont {B.}~\bibnamefont
  {{Ahmedov}}}, \bibinfo {author} {\bibfnamefont {O.}~\bibnamefont
  {{Rahimov}}}, \ and\ \bibinfo {author} {\bibfnamefont {U.}~\bibnamefont
  {{Salikhbaev}}},\ }\href {\doibase 10.1088/0031-8949/89/8/084008} {\bibfield
  {journal} {\bibinfo  {journal} {Physica Scripta}\ }\textbf {\bibinfo {volume}
  {89}},\ \bibinfo {eid} {084008} (\bibinfo {year} {2014})}\BibitemShut
  {NoStop}%
\bibitem [{\citenamefont {{Rahimov}}\ \emph {et~al.}(2011)\citenamefont
  {{Rahimov}}, \citenamefont {{Abdujabbarov}},\ and\ \citenamefont
  {{Ahmedov}}}]{Rahimov11a}%
  \BibitemOpen
  \bibfield  {author} {\bibinfo {author} {\bibfnamefont {O.~G.}\ \bibnamefont
  {{Rahimov}}}, \bibinfo {author} {\bibfnamefont {A.~A.}\ \bibnamefont
  {{Abdujabbarov}}}, \ and\ \bibinfo {author} {\bibfnamefont {B.~J.}\
  \bibnamefont {{Ahmedov}}},\ }\href {\doibase 10.1007/s10509-011-0755-1}
  {\bibfield  {journal} {\bibinfo  {journal} {Astrophysics and Space Science}\
  }\textbf {\bibinfo {volume} {335}},\ \bibinfo {pages} {499} (\bibinfo {year}
  {2011})},\ \Eprint {http://arxiv.org/abs/1105.4543} {arXiv:1105.4543
  [astro-ph.SR]} \BibitemShut {NoStop}%
\bibitem [{\citenamefont {{Rahimov}}(2011)}]{Rahimov11}%
  \BibitemOpen
  \bibfield  {author} {\bibinfo {author} {\bibfnamefont {O.~G.}\ \bibnamefont
  {{Rahimov}}},\ }\href {\doibase 10.1142/S0217732311034931} {\bibfield
  {journal} {\bibinfo  {journal} {Modern Physics Letters A}\ }\textbf {\bibinfo
  {volume} {26}},\ \bibinfo {pages} {399} (\bibinfo {year} {2011})},\ \Eprint
  {http://arxiv.org/abs/1012.1481} {arXiv:1012.1481 [gr-qc]} \BibitemShut
  {NoStop}%
\bibitem [{\citenamefont {Haydarov}\ \emph {et~al.}(2020)\citenamefont
  {Haydarov}, \citenamefont {Abdujabbarov}, \citenamefont {Rayimbaev},\ and\
  \citenamefont {Ahmedov}}]{Haydarov20}%
  \BibitemOpen
  \bibfield  {author} {\bibinfo {author} {\bibfnamefont {K.}~\bibnamefont
  {Haydarov}}, \bibinfo {author} {\bibfnamefont {A.}~\bibnamefont
  {Abdujabbarov}}, \bibinfo {author} {\bibfnamefont {J.}~\bibnamefont
  {Rayimbaev}}, \ and\ \bibinfo {author} {\bibfnamefont {B.}~\bibnamefont
  {Ahmedov}},\ }\href {\doibase 10.3390/universe6030044} {\bibfield  {journal}
  {\bibinfo  {journal} {Universe}\ }\textbf {\bibinfo {volume} {6}} (\bibinfo
  {year} {2020}),\ 10.3390/universe6030044}\BibitemShut {NoStop}%
\bibitem [{\citenamefont {{Abdujabbarov}}\ \emph {et~al.}(2020)\citenamefont
  {{Abdujabbarov}}, \citenamefont {{Rayimbaev}}, \citenamefont {{Turimov}},\
  and\ \citenamefont {{Atamurotov}}}]{Abdujabbarov2020PDU}%
  \BibitemOpen
  \bibfield  {author} {\bibinfo {author} {\bibfnamefont {A.}~\bibnamefont
  {{Abdujabbarov}}}, \bibinfo {author} {\bibfnamefont {J.}~\bibnamefont
  {{Rayimbaev}}}, \bibinfo {author} {\bibfnamefont {B.}~\bibnamefont
  {{Turimov}}}, \ and\ \bibinfo {author} {\bibfnamefont {F.}~\bibnamefont
  {{Atamurotov}}},\ }\href {\doibase 10.1016/j.dark.2020.100715} {\bibfield
  {journal} {\bibinfo  {journal} {Physics of the Dark Universe}\ }\textbf
  {\bibinfo {volume} {30}},\ \bibinfo {eid} {100715} (\bibinfo {year}
  {2020})}\BibitemShut {NoStop}%
\bibitem [{\citenamefont {{Narzilloev}}\ \emph {et~al.}(2020)\citenamefont
  {{Narzilloev}}, \citenamefont {{Rayimbaev}}, \citenamefont {{Shaymatov}},
  \citenamefont {{Abdujabbarov}}, \citenamefont {{Ahmedov}},\ and\
  \citenamefont {{Bambi}}}]{Narzilloev2020PhRvDstringy}%
  \BibitemOpen
  \bibfield  {author} {\bibinfo {author} {\bibfnamefont {B.}~\bibnamefont
  {{Narzilloev}}}, \bibinfo {author} {\bibfnamefont {J.}~\bibnamefont
  {{Rayimbaev}}}, \bibinfo {author} {\bibfnamefont {S.}~\bibnamefont
  {{Shaymatov}}}, \bibinfo {author} {\bibfnamefont {A.}~\bibnamefont
  {{Abdujabbarov}}}, \bibinfo {author} {\bibfnamefont {B.}~\bibnamefont
  {{Ahmedov}}}, \ and\ \bibinfo {author} {\bibfnamefont {C.}~\bibnamefont
  {{Bambi}}},\ }\href {\doibase 10.1103/PhysRevD.102.044013} {\bibfield
  {journal} {\bibinfo  {journal} {Physical Review D}\ }\textbf {\bibinfo
  {volume} {102}},\ \bibinfo {eid} {044013} (\bibinfo {year} {2020})},\ \Eprint
  {http://arxiv.org/abs/2007.12462} {arXiv:2007.12462 [gr-qc]} \BibitemShut
  {NoStop}%
\bibitem [{\citenamefont {{Rayimbaev}}\ \emph
  {et~al.}(2020{\natexlab{b}})\citenamefont {{Rayimbaev}}, \citenamefont
  {{Figueroa}}, \citenamefont {{Stuchl{\'\i}k}},\ and\ \citenamefont
  {{Juraev}}}]{Rayimbaev2020PhRvD}%
  \BibitemOpen
  \bibfield  {author} {\bibinfo {author} {\bibfnamefont {J.}~\bibnamefont
  {{Rayimbaev}}}, \bibinfo {author} {\bibfnamefont {M.}~\bibnamefont
  {{Figueroa}}}, \bibinfo {author} {\bibfnamefont {Z.}~\bibnamefont
  {{Stuchl{\'\i}k}}}, \ and\ \bibinfo {author} {\bibfnamefont {B.}~\bibnamefont
  {{Juraev}}},\ }\href {\doibase 10.1103/PhysRevD.101.104045} {\bibfield
  {journal} {\bibinfo  {journal} {Phys. Rev. D}\ }\textbf {\bibinfo {volume}
  {101}},\ \bibinfo {eid} {104045} (\bibinfo {year}
  {2020}{\natexlab{b}})}\BibitemShut {NoStop}%
\bibitem [{\citenamefont {Turimov}\ \emph {et~al.}(2020)\citenamefont
  {Turimov}, \citenamefont {Rayimbaev}, \citenamefont {Abdujabbarov},
  \citenamefont {Ahmedov},\ and\ \citenamefont
  {Stuchl\'{\i}k}}]{TurimovPhysRevD2020}%
  \BibitemOpen
  \bibfield  {author} {\bibinfo {author} {\bibfnamefont {B.}~\bibnamefont
  {Turimov}}, \bibinfo {author} {\bibfnamefont {J.}~\bibnamefont {Rayimbaev}},
  \bibinfo {author} {\bibfnamefont {A.}~\bibnamefont {Abdujabbarov}}, \bibinfo
  {author} {\bibfnamefont {B.}~\bibnamefont {Ahmedov}}, \ and\ \bibinfo
  {author} {\bibfnamefont {Z.~c.~v.}\ \bibnamefont {Stuchl\'{\i}k}},\ }\href
  {\doibase 10.1103/PhysRevD.102.064052} {\bibfield  {journal} {\bibinfo
  {journal} {Phys. Rev. D}\ }\textbf {\bibinfo {volume} {102}},\ \bibinfo
  {pages} {064052} (\bibinfo {year} {2020})}\BibitemShut {NoStop}%
\bibitem [{\citenamefont {{Morozova}}\ \emph {et~al.}(2014)\citenamefont
  {{Morozova}}, \citenamefont {{Rezzolla}},\ and\ \citenamefont
  {{Ahmedov}}}]{MorozovaV2014PhRvD}%
  \BibitemOpen
  \bibfield  {author} {\bibinfo {author} {\bibfnamefont {V.~S.}\ \bibnamefont
  {{Morozova}}}, \bibinfo {author} {\bibfnamefont {L.}~\bibnamefont
  {{Rezzolla}}}, \ and\ \bibinfo {author} {\bibfnamefont {B.~J.}\ \bibnamefont
  {{Ahmedov}}},\ }\href {\doibase 10.1103/PhysRevD.89.104030} {\bibfield
  {journal} {\bibinfo  {journal} {Phys.Rev.D}\ }\textbf {\bibinfo {volume}
  {89}},\ \bibinfo {eid} {104030} (\bibinfo {year} {2014})},\ \Eprint
  {http://arxiv.org/abs/1310.3575} {arXiv:1310.3575 [gr-qc]} \BibitemShut
  {NoStop}%
\bibitem [{\citenamefont {{Vrba}}\ \emph {et~al.}(2020)\citenamefont {{Vrba}},
  \citenamefont {{Abdujabbarov}}, \citenamefont {{Kolo{\v{s}}}}, \citenamefont
  {{Ahmedov}}, \citenamefont {{Stuchl{\'\i}k}},\ and\ \citenamefont
  {{Rayimbaev}}}]{Vrba2020PhRvD}%
  \BibitemOpen
  \bibfield  {author} {\bibinfo {author} {\bibfnamefont {J.}~\bibnamefont
  {{Vrba}}}, \bibinfo {author} {\bibfnamefont {A.}~\bibnamefont
  {{Abdujabbarov}}}, \bibinfo {author} {\bibfnamefont {M.}~\bibnamefont
  {{Kolo{\v{s}}}}}, \bibinfo {author} {\bibfnamefont {B.}~\bibnamefont
  {{Ahmedov}}}, \bibinfo {author} {\bibfnamefont {Z.}~\bibnamefont
  {{Stuchl{\'\i}k}}}, \ and\ \bibinfo {author} {\bibfnamefont {J.}~\bibnamefont
  {{Rayimbaev}}},\ }\href {\doibase 10.1103/PhysRevD.101.124039} {\bibfield
  {journal} {\bibinfo  {journal} {Phys.Rev.D}\ }\textbf {\bibinfo {volume}
  {101}},\ \bibinfo {eid} {124039} (\bibinfo {year} {2020})}\BibitemShut
  {NoStop}%
\bibitem [{\citenamefont {{Vrba}}\ \emph {et~al.}(2019)\citenamefont {{Vrba}},
  \citenamefont {{Abdujabbarov}}, \citenamefont {{Tursunov}}, \citenamefont
  {{Ahmedov}},\ and\ \citenamefont {{Stuchl{\'\i}k}}}]{Vrba2019EPJC}%
  \BibitemOpen
  \bibfield  {author} {\bibinfo {author} {\bibfnamefont {J.}~\bibnamefont
  {{Vrba}}}, \bibinfo {author} {\bibfnamefont {A.}~\bibnamefont
  {{Abdujabbarov}}}, \bibinfo {author} {\bibfnamefont {A.}~\bibnamefont
  {{Tursunov}}}, \bibinfo {author} {\bibfnamefont {B.}~\bibnamefont
  {{Ahmedov}}}, \ and\ \bibinfo {author} {\bibfnamefont {Z.}~\bibnamefont
  {{Stuchl{\'\i}k}}},\ }\href {\doibase 10.1140/epjc/s10052-019-7286-2}
  {\bibfield  {journal} {\bibinfo  {journal} {European Physical Journal C}\
  }\textbf {\bibinfo {volume} {79}},\ \bibinfo {eid} {778} (\bibinfo {year}
  {2019})},\ \Eprint {http://arxiv.org/abs/1909.12026} {arXiv:1909.12026
  [gr-qc]} \BibitemShut {NoStop}%
\bibitem [{\citenamefont {Akiyama}\ \emph {et~al.}(2019)\citenamefont {Akiyama}
  \emph {et~al.}}]{EventHorizonTelescope:2019dse}%
  \BibitemOpen
  \bibfield  {author} {\bibinfo {author} {\bibfnamefont {K.}~\bibnamefont
  {Akiyama}} \emph {et~al.} (\bibinfo {collaboration} {Event Horizon
  Telescope}),\ }\href {\doibase 10.3847/2041-8213/ab0ec7} {\bibfield
  {journal} {\bibinfo  {journal} {Astrophys. J. Lett.}\ }\textbf {\bibinfo
  {volume} {875}},\ \bibinfo {pages} {L1} (\bibinfo {year} {2019})},\ \Eprint
  {http://arxiv.org/abs/1906.11238} {arXiv:1906.11238 [astro-ph.GA]}
  \BibitemShut {NoStop}%
\bibitem [{\citenamefont {Ghasemi-Nodehi}\ \emph {et~al.}(2020)\citenamefont
  {Ghasemi-Nodehi}, \citenamefont {Azreg-A\"\i{}nou}, \citenamefont {Jusufi},\
  and\ \citenamefont {Jamil}}]{Ghasemi-Nodehi:2020oiz}%
  \BibitemOpen
  \bibfield  {author} {\bibinfo {author} {\bibfnamefont {M.}~\bibnamefont
  {Ghasemi-Nodehi}}, \bibinfo {author} {\bibfnamefont {M.}~\bibnamefont
  {Azreg-A\"\i{}nou}}, \bibinfo {author} {\bibfnamefont {K.}~\bibnamefont
  {Jusufi}}, \ and\ \bibinfo {author} {\bibfnamefont {M.}~\bibnamefont
  {Jamil}},\ }\href {\doibase 10.1103/PhysRevD.102.104032} {\bibfield
  {journal} {\bibinfo  {journal} {Phys. Rev. D}\ }\textbf {\bibinfo {volume}
  {102}},\ \bibinfo {pages} {104032} (\bibinfo {year} {2020})},\ \Eprint
  {http://arxiv.org/abs/2011.02276} {arXiv:2011.02276 [gr-qc]} \BibitemShut
  {NoStop}%
\bibitem [{\citenamefont {Jusufi}\ \emph {et~al.}(2021)\citenamefont {Jusufi},
  \citenamefont {Azreg-A\"\i{}nou}, \citenamefont {Jamil}, \citenamefont {Wei},
  \citenamefont {Wu},\ and\ \citenamefont {Wang}}]{Jusufi:2020odz}%
  \BibitemOpen
  \bibfield  {author} {\bibinfo {author} {\bibfnamefont {K.}~\bibnamefont
  {Jusufi}}, \bibinfo {author} {\bibfnamefont {M.}~\bibnamefont
  {Azreg-A\"\i{}nou}}, \bibinfo {author} {\bibfnamefont {M.}~\bibnamefont
  {Jamil}}, \bibinfo {author} {\bibfnamefont {S.-W.}\ \bibnamefont {Wei}},
  \bibinfo {author} {\bibfnamefont {Q.}~\bibnamefont {Wu}}, \ and\ \bibinfo
  {author} {\bibfnamefont {A.}~\bibnamefont {Wang}},\ }\href {\doibase
  10.1103/PhysRevD.103.024013} {\bibfield  {journal} {\bibinfo  {journal}
  {Phys. Rev. D}\ }\textbf {\bibinfo {volume} {103}},\ \bibinfo {pages}
  {024013} (\bibinfo {year} {2021})},\ \Eprint
  {http://arxiv.org/abs/2008.08450} {arXiv:2008.08450 [gr-qc]} \BibitemShut
  {NoStop}%
\bibitem [{\citenamefont {Liu}\ \emph {et~al.}(2020)\citenamefont {Liu},
  \citenamefont {Zhu}, \citenamefont {Wu}, \citenamefont {Jusufi},
  \citenamefont {Jamil}, \citenamefont {Azreg-A\"\i{}nou},\ and\ \citenamefont
  {Wang}}]{Liu:2020ola}%
  \BibitemOpen
  \bibfield  {author} {\bibinfo {author} {\bibfnamefont {C.}~\bibnamefont
  {Liu}}, \bibinfo {author} {\bibfnamefont {T.}~\bibnamefont {Zhu}}, \bibinfo
  {author} {\bibfnamefont {Q.}~\bibnamefont {Wu}}, \bibinfo {author}
  {\bibfnamefont {K.}~\bibnamefont {Jusufi}}, \bibinfo {author} {\bibfnamefont
  {M.}~\bibnamefont {Jamil}}, \bibinfo {author} {\bibfnamefont
  {M.}~\bibnamefont {Azreg-A\"\i{}nou}}, \ and\ \bibinfo {author}
  {\bibfnamefont {A.}~\bibnamefont {Wang}},\ }\href {\doibase
  10.1103/PhysRevD.101.084001} {\bibfield  {journal} {\bibinfo  {journal}
  {Phys. Rev. D}\ }\textbf {\bibinfo {volume} {101}},\ \bibinfo {pages}
  {084001} (\bibinfo {year} {2020})},\ \bibinfo {note} {[Erratum: Phys.Rev.D
  103, 089902 (2021)]},\ \Eprint {http://arxiv.org/abs/2003.00477}
  {arXiv:2003.00477 [gr-qc]} \BibitemShut {NoStop}%
\bibitem [{\citenamefont {Jusufi}\ \emph {et~al.}(2020)\citenamefont {Jusufi},
  \citenamefont {Jamil}, \citenamefont {Chakrabarty}, \citenamefont {Wu},
  \citenamefont {Bambi},\ and\ \citenamefont {Wang}}]{Jusufi:2019caq}%
  \BibitemOpen
  \bibfield  {author} {\bibinfo {author} {\bibfnamefont {K.}~\bibnamefont
  {Jusufi}}, \bibinfo {author} {\bibfnamefont {M.}~\bibnamefont {Jamil}},
  \bibinfo {author} {\bibfnamefont {H.}~\bibnamefont {Chakrabarty}}, \bibinfo
  {author} {\bibfnamefont {Q.}~\bibnamefont {Wu}}, \bibinfo {author}
  {\bibfnamefont {C.}~\bibnamefont {Bambi}}, \ and\ \bibinfo {author}
  {\bibfnamefont {A.}~\bibnamefont {Wang}},\ }\href {\doibase
  10.1103/PhysRevD.101.044035} {\bibfield  {journal} {\bibinfo  {journal}
  {Phys. Rev. D}\ }\textbf {\bibinfo {volume} {101}},\ \bibinfo {pages}
  {044035} (\bibinfo {year} {2020})},\ \Eprint
  {http://arxiv.org/abs/1911.07520} {arXiv:1911.07520 [gr-qc]} \BibitemShut
  {NoStop}%
\bibitem [{\citenamefont {{Afrin}}\ \emph {et~al.}(2021)\citenamefont
  {{Afrin}}, \citenamefont {{Kumar}},\ and\ \citenamefont
  {{Ghosh}}}]{Afrin2021a}%
  \BibitemOpen
  \bibfield  {author} {\bibinfo {author} {\bibfnamefont {M.}~\bibnamefont
  {{Afrin}}}, \bibinfo {author} {\bibfnamefont {R.}~\bibnamefont {{Kumar}}}, \
  and\ \bibinfo {author} {\bibfnamefont {S.~G.}\ \bibnamefont {{Ghosh}}},\
  }\href {\doibase 10.1093/mnras/stab1260} {\bibfield  {journal} {\bibinfo
  {journal} {Mon.~Not.~R.~Astron.~Soc.}\ }\textbf {\bibinfo {volume} {504}},\
  \bibinfo {pages} {5927} (\bibinfo {year} {2021})},\ \Eprint
  {http://arxiv.org/abs/2103.11417} {arXiv:2103.11417 [gr-qc]} \BibitemShut
  {NoStop}%
\bibitem [{\citenamefont {{Atamurotov}}\ \emph {et~al.}(2013)\citenamefont
  {{Atamurotov}}, \citenamefont {{Abdujabbarov}},\ and\ \citenamefont
  {{Ahmedov}}}]{Atamurotov2013a}%
  \BibitemOpen
  \bibfield  {author} {\bibinfo {author} {\bibfnamefont {F.}~\bibnamefont
  {{Atamurotov}}}, \bibinfo {author} {\bibfnamefont {A.}~\bibnamefont
  {{Abdujabbarov}}}, \ and\ \bibinfo {author} {\bibfnamefont {B.}~\bibnamefont
  {{Ahmedov}}},\ }\href {\doibase 10.1103/PhysRevD.88.064004} {\bibfield
  {journal} {\bibinfo  {journal} {Phys. Rev. D}\ }\textbf {\bibinfo {volume}
  {88}},\ \bibinfo {eid} {064004} (\bibinfo {year} {2013})}\BibitemShut
  {NoStop}%
\bibitem [{\citenamefont {Bambi}(2015)}]{Bambi:2014mla}%
  \BibitemOpen
  \bibfield  {author} {\bibinfo {author} {\bibfnamefont {C.}~\bibnamefont
  {Bambi}},\ }\href {\doibase 10.1088/0264-9381/32/6/065005} {\bibfield
  {journal} {\bibinfo  {journal} {Class. Quant. Grav.}\ }\textbf {\bibinfo
  {volume} {32}},\ \bibinfo {pages} {065005} (\bibinfo {year} {2015})},\
  \Eprint {http://arxiv.org/abs/1409.0310} {arXiv:1409.0310 [gr-qc]}
  \BibitemShut {NoStop}%
\bibitem [{\citenamefont {{Abdujabbarov}}\ \emph {et~al.}(2013)\citenamefont
  {{Abdujabbarov}}, \citenamefont {{Atamurotov}}, \citenamefont {{Kucukakca}},
  \citenamefont {{Ahmedov}},\ and\ \citenamefont
  {{Camci}}}]{Abdujabbarov2013a}%
  \BibitemOpen
  \bibfield  {author} {\bibinfo {author} {\bibfnamefont {A.}~\bibnamefont
  {{Abdujabbarov}}}, \bibinfo {author} {\bibfnamefont {F.}~\bibnamefont
  {{Atamurotov}}}, \bibinfo {author} {\bibfnamefont {Y.}~\bibnamefont
  {{Kucukakca}}}, \bibinfo {author} {\bibfnamefont {B.}~\bibnamefont
  {{Ahmedov}}}, \ and\ \bibinfo {author} {\bibfnamefont {U.}~\bibnamefont
  {{Camci}}},\ }\href {\doibase 10.1007/s10509-012-1337-6} {\bibfield
  {journal} {\bibinfo  {journal} {Astrophys. Space. Sci.}\ }\textbf {\bibinfo
  {volume} {344}},\ \bibinfo {pages} {429} (\bibinfo {year} {2013})},\ \Eprint
  {http://arxiv.org/abs/1212.4949} {arXiv:1212.4949 [physics.gen-ph]}
  \BibitemShut {NoStop}%
\bibitem [{\citenamefont {Atamurotov}\ \emph {et~al.}(2016)\citenamefont
  {Atamurotov}, \citenamefont {G.~Ghosh},\ and\ \citenamefont
  {Ahmedov}}]{Far:2016c}%
  \BibitemOpen
  \bibfield  {author} {\bibinfo {author} {\bibfnamefont {F.}~\bibnamefont
  {Atamurotov}}, \bibinfo {author} {\bibfnamefont {S.}~\bibnamefont
  {G.~Ghosh}}, \ and\ \bibinfo {author} {\bibfnamefont {B.}~\bibnamefont
  {Ahmedov}},\ }\href {\doibase 10.1140/epjc/s10052-016-4122-9} {\bibfield
  {journal} {\bibinfo  {journal} {Eur.~Phys.~J.~C.}\ }\textbf {\bibinfo
  {volume} {76}},\ \bibinfo {pages} {273} (\bibinfo {year} {2016})}\BibitemShut
  {NoStop}%
\bibitem [{\citenamefont {{Sharif}}\ and\ \citenamefont
  {{Shahzadi}}(2018)}]{Sharif2018JETP}%
  \BibitemOpen
  \bibfield  {author} {\bibinfo {author} {\bibfnamefont {M.}~\bibnamefont
  {{Sharif}}}\ and\ \bibinfo {author} {\bibfnamefont {M.}~\bibnamefont
  {{Shahzadi}}},\ }\href {\doibase 10.1134/S1063776118090182} {\bibfield
  {journal} {\bibinfo  {journal} {Soviet Journal of Experimental and
  Theoretical Physics}\ }\textbf {\bibinfo {volume} {127}},\ \bibinfo {pages}
  {491} (\bibinfo {year} {2018})}\BibitemShut {NoStop}%
\bibitem [{\citenamefont {{Moffat}}\ and\ \citenamefont
  {{Rahvar}}(2013{\natexlab{b}})}]{Moffat2013MNRAS}%
  \BibitemOpen
  \bibfield  {author} {\bibinfo {author} {\bibfnamefont {J.~W.}\ \bibnamefont
  {{Moffat}}}\ and\ \bibinfo {author} {\bibfnamefont {S.}~\bibnamefont
  {{Rahvar}}},\ }\href {\doibase 10.1093/mnras/stt1670} {\bibfield  {journal}
  {\bibinfo  {journal} {Mon.Not.R.Astron.}\ }\textbf {\bibinfo {volume}
  {436}},\ \bibinfo {pages} {1439} (\bibinfo {year} {2013}{\natexlab{b}})},\
  \Eprint {http://arxiv.org/abs/1306.6383} {arXiv:1306.6383 [astro-ph.GA]}
  \BibitemShut {NoStop}%
\bibitem [{\citenamefont {{Thorne}}(1974)}]{Thorne1974ApJ}%
  \BibitemOpen
  \bibfield  {author} {\bibinfo {author} {\bibfnamefont {K.~S.}\ \bibnamefont
  {{Thorne}}},\ }\href {\doibase 10.1086/152991} {\bibfield  {journal}
  {\bibinfo  {journal} {Astrophys. Jour.}\ }\textbf {\bibinfo {volume} {191}},\
  \bibinfo {pages} {507} (\bibinfo {year} {1974})}\BibitemShut {NoStop}%
\bibitem [{\citenamefont {{Kolo{\v s}}}\ \emph {et~al.}(2015)\citenamefont
  {{Kolo{\v s}}}, \citenamefont {{Stuchl{\'{\i}}k}},\ and\ \citenamefont
  {{Tursunov}}}]{Kolos15}%
  \BibitemOpen
  \bibfield  {author} {\bibinfo {author} {\bibfnamefont {M.}~\bibnamefont
  {{Kolo{\v s}}}}, \bibinfo {author} {\bibfnamefont {Z.}~\bibnamefont
  {{Stuchl{\'{\i}}k}}}, \ and\ \bibinfo {author} {\bibfnamefont
  {A.}~\bibnamefont {{Tursunov}}},\ }\href {\doibase
  10.1088/0264-9381/32/16/165009} {\bibfield  {journal} {\bibinfo  {journal}
  {Classical and Quantum Gravity}\ }\textbf {\bibinfo {volume} {32}},\ \bibinfo
  {eid} {165009} (\bibinfo {year} {2015})},\ \Eprint
  {http://arxiv.org/abs/1506.06799} {arXiv:1506.06799 [gr-qc]} \BibitemShut
  {NoStop}%
\end{thebibliography}%
\end{document}